\documentclass[final,3p,times,authoryear]{elsarticle}

\usepackage{graphicx}%
\usepackage{multirow}%
\usepackage{amsmath,amssymb,amsfonts}%
\usepackage{amsthm}%
\usepackage{mathrsfs}%
\usepackage[title]{appendix}%
\usepackage[table]{xcolor}%
\usepackage{textcomp}%
\usepackage{manyfoot}%
\usepackage{booktabs}%
\usepackage{algorithm}%
\usepackage{algorithmicx}%
\usepackage{algpseudocode}%
\usepackage{listings}%
\usepackage{subfig}
\usepackage{listings}
\usepackage{hyperref}
\usepackage{comment}
\usepackage{rotating}
\usepackage{makecell}
\usepackage[most]{tcolorbox}
\usepackage{array}

\definecolor{lightgray}{gray}{0.9}
\definecolor{mygreenish}{HTML}{A0CCBC}

\tcbset{
  mynote/.style={
    colback=mygreenish!70!white,
    colframe=black!40,
    coltitle=black,
    boxrule=0.8pt,
    arc=5mm,
    top=3pt,
    bottom=3pt,
    left=5pt,
    right=5pt,
    fonttitle=\bfseries,
    enhanced,
    attach boxed title to top left={xshift=1em,yshift=-1mm},
    boxed title style={colback=mygreenish!70!white,boxrule=0pt},
  }
}

\journal{Computers \& Education}

\begin{document}

\begin{frontmatter}

\title{Automatic Mind Wandering Detection in Educational Settings: A Systematic Review and Multimodal Benchmarking\\ 
\textit{\small A Preprint}} 

\author[1,2]{Anna Bodonhelyi\corref{cor1}}
\author[1]{Augustin Curinier}
\author[1,2]{Babette Bühler}
\author[3]{Gerrit Anders}
\author[3,4]{Lisa Rausch}
\author[3,4]{Markus Huff}
\author[5]{Ulrich Trautwein}
\author[6,7]{Ralph Ewerth}
\author[3]{Peter Gerjets}
\author[1]{Enkelejda Kasneci}

\affiliation[1]{organization={Chair for Human-Centered Technologies for Learning, Technical University of Munich},
            addressline={Arcisstr. 21.}, 
            city={Munich},
            postcode={80333}, 
            country={Germany}}
\affiliation[2]{organization={Munich Center for Machine Learning (MCML)}}    
\affiliation[3]{organization={Multimodal Interaction Lab, Leibniz-Institut für Wissensmedien},
             addressline={Schleichstr. 6}, 
             city={Tübingen},
             postcode={72076}, 
             country={Germany}}             
\affiliation[4]{organization={Faculty of Science, Applied Cognitive Psychology, Eberhard Karls University of Tübingen},
             addressline={Schleichstr. 4}, 
             city={Tübingen},
             postcode={72076}, 
             country={Germany}} 
\affiliation[5]{organization={Hector Research Institute of Education Sciences and Psychology, Eberhard Karls University of Tübingen},
             addressline={Europastr. 6}, 
             city={Tübingen},
             postcode={72072}, 
             country={Germany}} 
\affiliation[6]{organization={Chair of Multimodal Modelling and Machine Learning, Marburg University \& Hessian Center for Artificial Intelligence (hessian.AI)},
             addressline={Hans-Meerwein-Str. 6}, 
             city={Marburg},
             postcode={35032}, 
             country={Germany}}
\affiliation[7]{organization={TIB – Leibniz Information Centre for Science and Technology},
             addressline={Welfengarten 1 B}, 
             city={Hannover},
             postcode={30167}, 
             country={Germany}}

\cortext[cor1]{Corresponding author: Anna Bodonhelyi, anna.bodonhelyi@tum.de}

\begin{abstract}
Detecting mind wandering is crucial in online education, and it occurs 30\% of the time, as it directly impacts learners’ retention, comprehension, and overall success in self-directed learning environments. Integrating automated detection algorithms enables the deployment of targeted interventions within adaptive learning environments, paving the way for more responsive and personalized educational systems. However, progress is hampered by a lack of coherent frameworks for identifying mind wandering in online environments. This work presents a comprehensive systematic review and benchmark of mind wandering detection across 14 datasets covering EEG, facial video, eye tracking, and physiological signals in educational settings, motivated by the challenges in achieving reliable detection and the inconsistency of results across studies caused by variations in models, preprocessing approaches, and evaluation metrics. We implemented a generalizable preprocessing and feature extraction pipeline tailored to each modality, ensuring fair comparison across diverse experimental paradigms. 13 traditional machine learning and neural network models, including federated learning approaches, were evaluated on each dataset. In a novel ablation study, we explored mind wandering detection from post-probe data, motivated by findings that learners often re-engage with material after mind wandering episodes through re-reading or re-watching. Results highlight the potential and limitations of different modalities and classifiers for mind wandering detection, and point to new opportunities for supporting online learning. All code and preprocessing scripts are made openly available to support reproducibility and future research.
\end{abstract}


\end{frontmatter}

\insert\footins{\noindent\footnotesize Preprint submitted to Computers \& Education}

\lstset{basicstyle=\scriptsize}

 \section{Introduction}
Self-directed online learning, facilitated by platforms such as MOOCs and institutional environments \citep{fini2009technological, liyanagunawardena2015massive}, increasingly takes place in informal and distributed settings where sustaining attention is particularly challenging \citep{el2021did, simamora2020challenges, bylieva2021self}. In the absence of direct instructor support, learners are especially susceptible to distractions and mind wandering \citep{hollis2016mind, wammes2019disengagement}, a cognitive state in which attention shifts from the primary task to internal, task-unrelated thoughts \citep{smallwood2006restless}. This phenomenon is highly prevalent: people spend nearly half of their waking hours mind wandering (46.9\%), with task-unrelated thoughts occurring in at least 30\% of sampled moments across everyday activities \citep{killingsworth2010wandering}, while in learning contexts, learners mind wander during roughly 30\% of study time regardless of the learning situation \citep{wong2022task}. Similar patterns emerge in other attention-critical settings, with mind wandering reported during approximately 39–42\% of driving time \citep{yanko2014driving}. Recent findings indicate that nearly 50\% of adolescents experience such attentional shifts frequently in daily life \citep{zhang2025detrimental, pachai2016mind}. These episodes are often driven by environmental stressors and misalignment between cognitive demands and context, with factors such as childhood adversity or peer conflict triggering nonlinear increases in task-unrelated thoughts \citep{zhang2025detrimental, andrewes2019role}. Overall, mind wandering accounts for approximately 7\% of the variance in learning outcomes and is consistently associated with poorer performance across subjects and age groups.

Because mind wandering substantially impairs knowledge retention \citep{wong2022task}, there is growing interest in applying Machine Learning (ML) to detect it in real time, enabling adaptive learner support and data-driven insights for improving instructional design and digital learning outcomes. However, progress toward such applications is limited by the scarcity of publicly available datasets that are both multimodal and suitable for real-world deployment. Many existing datasets depend on costly and intrusive sensors, such as EEG headsets~\citep{china2023eeg, jin2020distinguishing} or dedicated eye-tracking systems~\citep{bixler2015, MMSART2022eeg}, which require controlled environments and hinder scalability. In contrast, webcam-based approaches~\citep{bosch2019automatic, lee2022predicting} offer a more accessible and practical alternative for large-scale or at-home use, albeit typically at the cost of lower detection accuracy.

Mind wandering research has advanced along two major disciplinary trajectories. In \textit{educational psychology}, it is typically examined through subjective measures such as self-reports collected via experience sampling methods (e.g., probe-caught or self-caught reports) or retrospective questionnaires. In contrast, \textit{computer science} approaches emphasize the automated, non-intrusive detection of mind wandering through behavioral and physiological signals, often coupled with ML techniques. These computational methods aim to enable large-scale, unobtrusive assessment, using self-reports primarily during the initial stages to establish ground truth labels for model training.

\begin{tcolorbox}[mynote,title={\textbf{Research gap}}]
There is a critical lack of systematic benchmarking across accessible sensing modalities and real-world learning contexts, which limits our understanding of which detection methods are most effective and scalable for mind wandering interventions.
\end{tcolorbox}

Bridging these fields requires a \textbf{solid methodological foundation} that integrates computational rigor with educational relevance. this work makes the following key contributions:
\begin{itemize}
    \item An \textbf{up-to-date systematic review} of automatic mind wandering detection research within educational contexts.
    \item Introduction of a new \textbf{eye-tracking} dataset (available on \href{https://zenodo.org/records/14097753?preview=1&token=eyJhbGciOiJIUzUxMiJ9.eyJpZCI6IjAyM2QxNWY1LWJlY2ItNDQ2ZC04YjNhLWUwYjdkZGU0MzU4ZCIsImRhdGEiOnt9LCJyYW5kb20iOiJkNTkyZmI0MzMxMWRiODVjNzJkNzhkNGYyOWEzMjMwYiJ9.MRLZkGqYIwafNVI-D9z1de1qY4o-iQ1fPJFK5NXgImrL095j3cWAoMhwjJHlbD3ZdeVWrmDuPH_ItHQY-wGMrQ}{Zenodo}).
    \item A \textbf{comprehensive benchmarking study} with a published benchmarking framework (available on \href{https://gitlab.lrz.de/hctl/mw-benchmarking/}{GitLab}) of widely used machine learning algorithms across 14 datasets.
    \item \textbf{Evidence-based recommendations} identifying the most effective machine learning methods for specific data modalities used in mind wandering detection.
    \item An ablation study introducing and evaluating a \textbf{novel sampling approach for data collection}, highlighting its impact on model performance compared to existing dataset conventions.
\end{itemize}

\section{Related Work}

\subsection{Existing Reviews on Understanding Mind Wandering}
Mind wandering has been examined from multiple disciplinary perspectives, and several reviews have summarized prior work in this area. Among these, the review by \citet{kuvar2023detecting} is the most relevant to the present study, as it focuses explicitly on automated, machine learning–based mind wandering detection. Synthesizing 42 studies, it introduces a structured framework encompassing six key dimensions: task type, ground truth collection, modality, machine learning approach, validation strategy, and evaluation metrics. However, the review is no longer fully up to date, includes studies from task domains beyond educational contexts, and reflects considerable heterogeneity in evaluation and validation practices, which limits cross-study comparability. While their review offers a multidisciplinary overview of existing systematic reviews on mind wandering, in the following we synthesize the most pertinent literature from the perspectives of educational psychology and computer science.

\paragraph{\textbf{Education}} More recent work has shifted toward understanding mind wandering in educational settings. \citet{dmello2021} conducted an interdisciplinary review highlighting how mind wandering impairs reading comprehension. 
A complementary meta-analysis by \citet{meziere_eye-movement_2025} focused on eye-movement correlates of mind wandering in 16 studies, revealing behavioral patterns. 
\citet{wong2022} expanded this scope through a meta-analysis of 130 samples, demonstrating that task-unrelated thoughts are prevalent in learning contexts and consistently associated with poorer outcomes. These reviews highlight the significance of mind wandering, but largely concentrate on its consequences rather than its detection.

\paragraph{\textbf{Cognitive Science}} The study of mind wandering has long attracted attention in cognitive science, with early reviews focusing on its theoretical foundations and underlying neural mechanisms. For example, \citet{smallwood_mind-wandering_2011} emphasized the role of decoupled processing during reading, where external attention fades in favor of internal thought. \citet{fox2015} conducted a meta-analysis of 24 neuroimaging studies, finding frequent activation in the default mode network during spontaneous thought, though other brain regions also contributed. Further exploring the content of mind wandering, \citet{banks_examining_2016} re-analyzed three earlier studies~\citep{banks_wheres_2014, banks_protective_2015, banks_understanding_2017} to examine the role of emotional valence, finding that both neutral and negative thoughts impair working memory. Similarly, \citet{andrews-hanna_neuroscience_2017} reframed spontaneous thought as a dynamic cognitive process encompassing mind wandering, creativity, and daydreaming, integrating insights from psychology, philosophy, and neuroscience.

\begin{tcolorbox}[mynote,title={\textbf{Domain-Specific Focus}}]
Our review provides an up-to-date, education-centered synthesis of mind wandering detection research, addressing prior limitations by focusing exclusively on learning contexts and applying standardized evaluation metrics for reliable cross-study comparisons.
\end{tcolorbox}

\subsection{Methodological Approaches for Mind Wandering Detection}
In the context of mind wandering research, validation of detection still fundamentally relies on self-reports, as there is currently no objective and reliable behavioral or neurophysiological marker available to identify mind wandering with certainty \citep{buhler2024lab, smallwood2015science}. Two widely used methods to capture mind wandering are probe-caught and self-caught approaches, where learners either respond to periodic attention probes \citep{smallwood2006restless} or voluntarily report noticed attentional lapses \citep{schooler2011meta}.
However, in learning environments, these methods pose challenges: frequent probes may disrupt the learning flow, while self-caught reports rely heavily on learners' metacognitive awareness. To address this diversity, we summarize commonly used sensing modalities and ground truthing strategies in \autoref{tab:modality_gt_summary}, based on our systematic review.

\paragraph{\textbf{Eye-Tracking}} ML based mind wandering detection increasingly relies on eye-tracking as the most common modality. Features such as fixation duration, pupil dilation, and saccade frequency are consistently linked to mind wandering \citep{kuvar2023review, faber2020}, though performance can be affected by environmental conditions and device quality~\citep{kasneci2024introduction}.

\paragraph{\textbf{Physiological Measures}} They provide more direct insight into cognitive and affective states. \textbf{Electroencephalogram (EEG)} recordings can capture neural patterns associated with mind wandering \citep{kawashima2017, china2023eeg, dong2021detection, jin2020distinguishing, dhindsa2019individualized, jin2019predicting}. Common features include band-specific power changes, with delta (1–4 Hz) and theta (4–8 Hz) activity linked to mind wandering, while alpha (8–13 Hz) and beta (13–30 Hz) bands also show predictive value \citep{jin2020distinguishing}. 
Despite their precision, EEG systems remain costly and intrusive, limiting their practicality for everyday educational settings. Alternatively, autonomic signals, such as \textbf{heart rate variability} \citep{buhler_detecting_2024}, \textbf{electrodermal activity (EDA)} \citep{brishtel2020}, and \textbf{photoplethysmography (PPG)} \citep{MMSART2022eeg} can be recorded using wearable devices, offering a more scalable solution for classroom or home-based settings. These measures, however, can be susceptible to motion artifacts and individual physiological variability.

\paragraph{\textbf{Facial Video}} It utilizes computer vision to interpret changes in facial expressions and head movements as proxies for mind wandering \citep{stewart2017, colorado2021, lee_predicting_2022, buhler_detecting_2024,buhler_temporal_2025,buhler2024task}. Previous datasets have commonly relied on extracted features such as OpenFace~\citep{baltrusaitis2018openface} for facial landmark detection, head pose estimation, facial action unit recognition, and eye-gaze estimation, as well as AffectNet~\citep{mollahosseini2017affectnet} or EmoNet~\citep{toisoul2021estimation} for deriving hidden emotion-related features. 
Although facial video benefits from the ubiquity of webcams in online education, its effectiveness may be limited by cultural and individual differences in expressiveness, as well as privacy concerns~\citep{goldberg2021attentive,sumer2021multimodal,FUTTERER2025100483}. 

\begin{tcolorbox}[mynote,title={\textbf{Unified Benchmarking}}]
We present a unified benchmarking framework that enables consistent comparison of mind wandering detection models across diverse datasets and modalities using standardized evaluation protocols.
\end{tcolorbox}

\section{Systematic Review}
Our work is guided by the following research questions: 
(1) What is the relationship between mind wandering and learning outcomes, in educational settings? 
(2) What methodological advances characterize recent approaches to mind wandering detection? (3) What are the current characteristics in datasets used for mind wandering research, and how do they influence detection performance and generalizability?.

\subsection{Methodology}
The systematic search was conducted on April 8, 2025, utilizing the \href{https://www.scopus.com}{Scopus} and \href{https://www.webofscience.com}{Web of Science} databases. The search string did not contain any date limitations to ensure comprehensive coverage of relevant work. For both databases, the search targeted the title, abstract, and keywords fields. The search strategy incorporated three primary groups of terms: \textbf{mind wandering} and its synonyms (e.g., task-unrelated thought, off-task thought), \textbf{automatic detection algorithms} (e.g., machine learning, deep neural network), and \textbf{modalities} (e.g., EEG, video). These main categories were combined using the logical ``AND'' operator to ensure that all essential aspects were covered, while terms within each category were linked using the ``OR'' operator to maximize inclusivity. The complete search string is detailed in \autoref{lst:search_query}.

\begin{figure}[b]
    \centering
    \includegraphics[width=13cm]{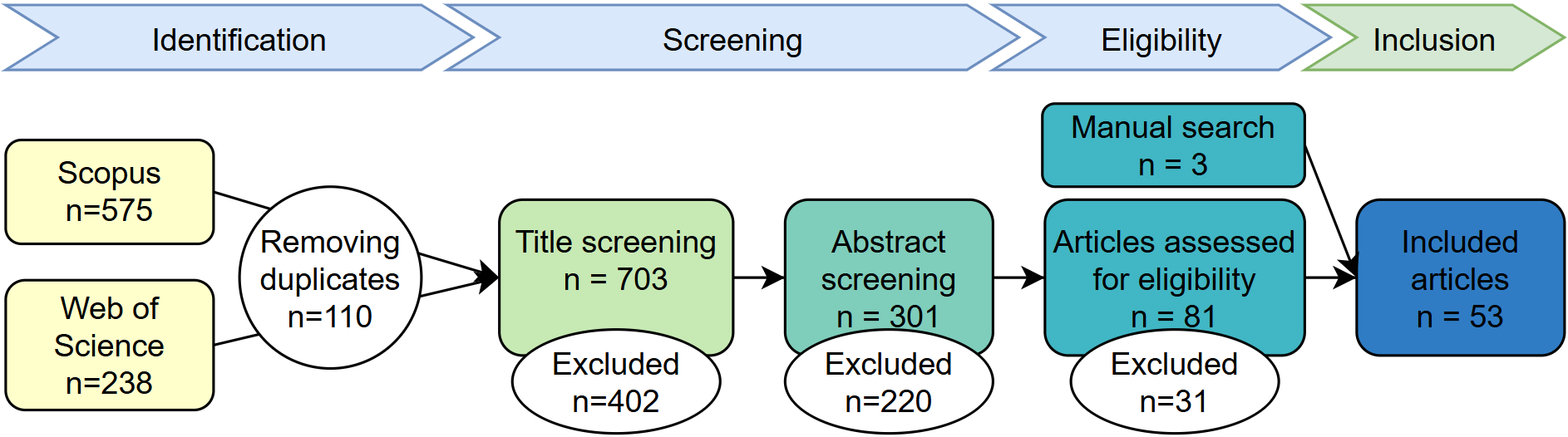}
    \caption{PRISMA~\citep{moher2009preferred} diagram for systematic search.}
    \label{fig:prisma}
\end{figure}

The review process adhered to the Preferred Reporting Items for Systematic Reviews and Meta-Analyses (PRISMA) guidelines~\citep{moher2009preferred} (\autoref{fig:prisma}). The initial search yielded 575 articles from Scopus and 238 from Web of Science, with 110 duplicates identified. The metadata of the remaining 703 unique articles was exported to a CSV file for further processing. Title and abstract screening was conducted using the Active Learning for Systematic Reviews (ASReview) software~\citep{van2020asreview} from the corresponding project~\citep{van2021open}. This screening process led to the exclusion of 402 articles following the title screening and an additional 220 articles after the abstract screening. A subsequent full-text screening eliminated an additional 31 articles. We included only studies on mind wandering detection in educational contexts, excluding driving- and clinical-focused work. Ultimately, 50 papers met the inclusion criteria and were included in this systematic analysis, with three additional studies identified through manual search based on related topics and citation graph exploration.

The included studies reported heterogeneous performance metrics (e.g., accuracy, $F_1$ score) under varying experimental settings. As such, the results were not directly comparable; however, they provided a broad empirical basis for cross-study comparison, which we later refer to in the benchmarking. The studies were grouped by modality and data collection method (\autoref{tab:modality_gt_summary}), ML algorithms (\autoref{tab:review_ml}), and modalities, environments and ML algorithms (\autoref{fig:rev_sum}). Data were manually curated, and manuscripts with missing or incomplete statistics were not included. Given the methodological heterogeneity of included studies, a meta-analysis was not performed; instead, a qualitative synthesis and comparative benchmarking were used to highlight consistent patterns and differences.

\subsection{Findings from the Systematic Review}
\begin{figure}[t]
    \centering
    \includegraphics[width=13cm]{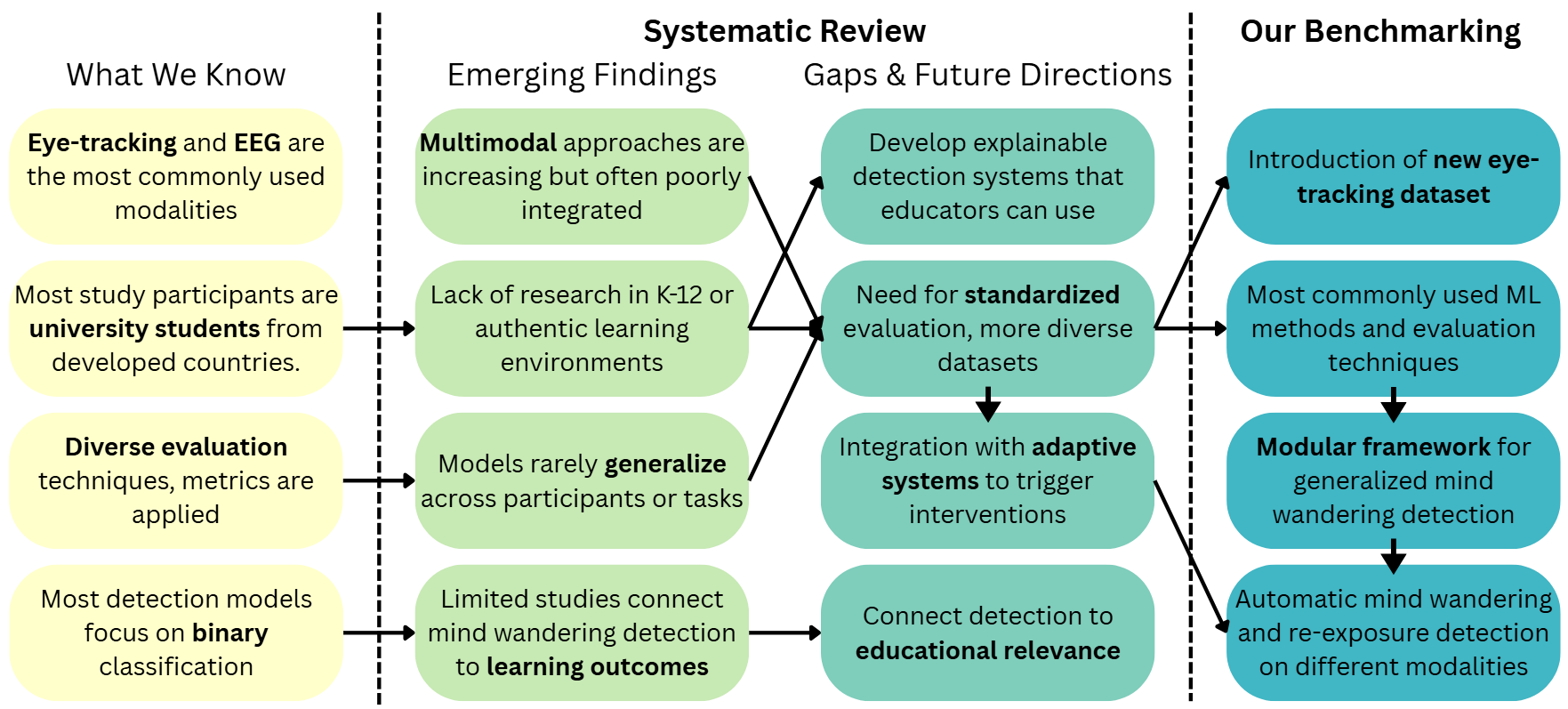}
    \caption{Summary of our findings in the systematic review with research gaps that created the need for our modular benchmarking framework.}
    \label{fig:rev_results}
\end{figure}

We reviewed literature on mind wandering detection in educational contexts and systematically analyzed the collected studies to evaluate their contributions. We summarize our findings in \autoref{fig:rev_results}. We identified research gaps through the review, which formed the foundation for our modular benchmarking framework.

\subsubsection{Automated Mind Wandering Detection Approaches in the Educational Domain}
\begin{figure}[t]
    \centering
    \includegraphics[width=10cm]{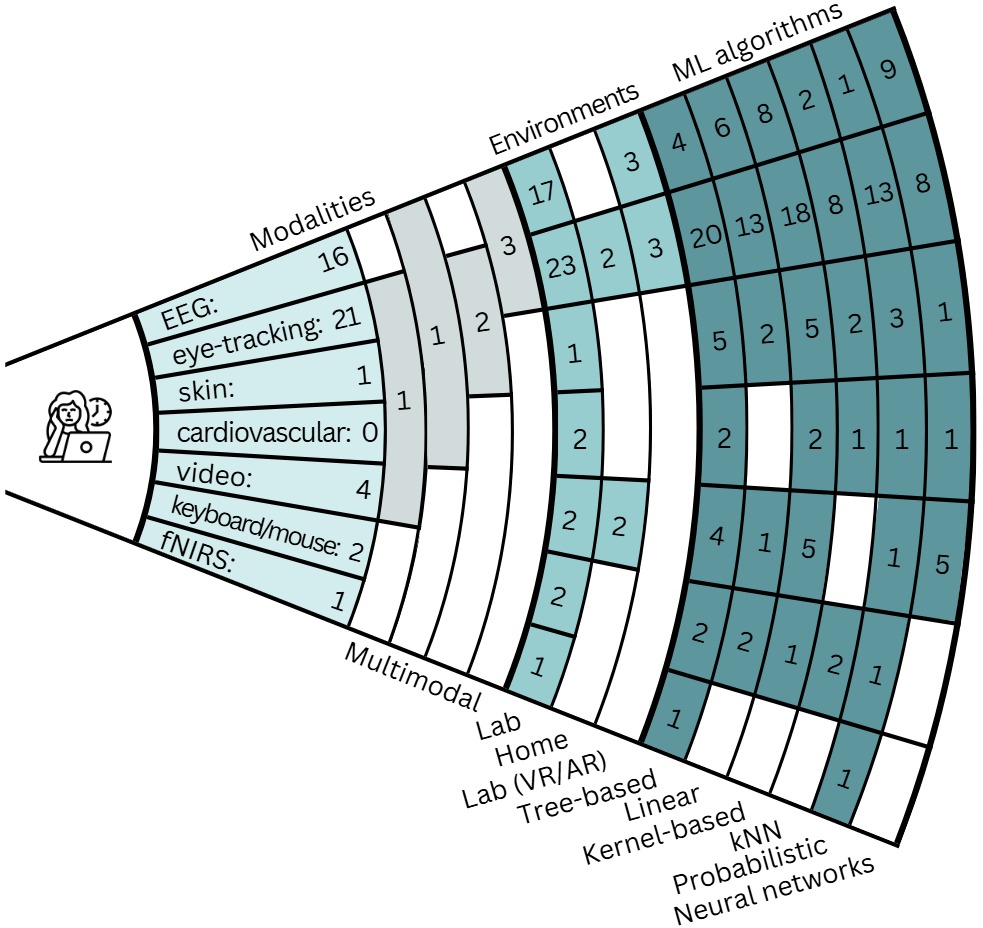}
    \caption{Review summary in three categories. Skin-based signals refer to physiological measurements such as EDA, skin conductance, and temperature, while cardiovascular signals include heart rate and PPG.}
    \label{fig:rev_sum}
\end{figure}

To characterize recent advances in mind wandering detection in educational contexts, we analyzed selected studies by modality, data collection environment, and applied ML algorithms (\autoref{fig:rev_sum}). EEG and eye tracking were most frequently used due to their high temporal resolution and established role in attention research. Seven datasets were multimodal~\citep{riby_elevated_2025, asish_classification_2024, long_multimodal_2024, buhler_detecting_2024, MMSART2022eeg, brishtel2020, bixler2015}, combining multiple signals to enhance detection robustness. Most data were collected in laboratory settings, where sensor use requires expert handling, while home-based studies relied on low-threshold, webcam-based methods~\citep{jaiyeola_one_2025, lee2022predicting}, highlighting the trade-off between ecological validity and data quality.

\paragraph{\textbf{Learning Outcomes}}
Although many studies have examined mind wandering detection in educational settings, few have applied machine learning to predict its impact on learning and comprehension, despite its well-documented negative effects~\citep{wong2022task}. For instance, one recent study~\citep{jaiyeola_one_2025} employed webcam-based eye-tracking during reading tasks and evaluated learning outcomes through multiple-choice questions, offering valuable insights into how mind wandering patterns correspond with learning performance.
In a related study, \citet{buhler_temporal_2025} used eye-tracking features and linear regression to detect mind wandering and predict comprehension in a lecture-viewing task via multiple-choice and open-ended questions. Frequent mind wandering correlated negatively with learning performance, underscoring its impact on effective learning.
Similarly, \citet{mills_eye-mind_2021} found that learners receiving a mind wandering intervention during video lectures showed improved delayed comprehension, highlighting eye-tracking’s potential to enhance learning. However, systematic links between detection accuracy and comprehension across tasks, modalities, and learner profiles remain underexplored.

\paragraph{\textbf{ML Algorithms}}
To examine algorithmic trends in mind wandering detection, we categorized models into six groups (\autoref{tab:review_ml}). Tree-based and kernel-based methods dominate due to their robustness, while linear and probabilistic models are now mainly used as baselines. Neural networks have recently gained prominence for their ability to model complex, multimodal, and sequential data, marking a shift toward more advanced architectures suited for large-scale applications. Instance-based methods remain rare due to scalability issues. 
Instance-based methods remain rare, primarily due to scalability concerns. Overall, the field is shifting from traditional models toward neural approaches suited for complex, large-scale data.

\begin{table}[t]
\centering
\footnotesize
\caption{Categorization of machine learning algorithms used in mind wandering detection studies in the educational domain.  If a study employed multiple algorithms, each instance was included individually to reflect methodological diversity.}
\begin{tabular}{m{0.8cm} m{1.7cm} m{10cm}}
\toprule
\textbf{Category} & \textbf{Algorithms} & \textbf{Published Works}\\
\midrule
Tree-based & Random Forest, XGBoost, Decision Tree, AdaBoost, GradBoost & \cite{jaiyeola_one_2025, buhler_temporal_2025, asish_classification_2024, buhler_detecting_2024, buhler2024lab, hutt2024webcam, rahnuma2024gazebased, kuvar2023keystroke, china2023eeg, asish_detecting_2022, khan2022execute, lee_predicting_2022, khan2022feature, MMSART2022eeg, liu2021fnirsbased, singha2021gaze, bixler2021crossed, chang2021efficient, dias_da_silva_wandering_2020, brishtel2020, tasika2020framework, gwizdka2019exploring, faber2018automated, stewart2017, hutt2016eyes, bixler2016automatic, mills2016automatic, bixler2015, bixler2015automatic, bixler2014toward} \\
\rowcolor[gray]{0.9} 
Linear & Logistic Regression, Linear Discriminant Analysis & \cite{chiossi_designing_2025, long_multimodal_2024, hutt2024webcam, china2023eeg, khan2022execute, asish_detecting_2022, khan2022feature, singha2021gaze, dong2021detection, dias_da_silva_wandering_2020, brishtel2020, vortmann_eeg-based_2019, faber2018automated, aliakbaryhosseinabadi_classification_2017, stewart2017, hutt2016eyes, bixler2016automatic, mills2016automatic, bixler2015, bixler2015automatic} \\
Kernel-based & SVMs, Gaussian Processes & \cite{buhler_temporal_2025, jaiyeola_one_2025, buhler_detecting_2024, buhler2024lab, hutt2024webcam, kuvar2023keystroke, china2023eeg, khan2022execute, MMSART2022eeg, lee_predicting_2022, mills_eye-mind_2021, colorado2021, singha2021gaze, bixler2021crossed, chang2021efficient, dong2021detection, brishtel2020, jin2020distinguishing, tasika2020framework, dhindsa2019individualized, jin2019predicting, stewart2017, hutt2016eyes, bixler2015, bixler2015automatic, bixler2014toward, grandchamp2014oculometric}\\
\rowcolor[gray]{0.9} 
Instance-based & K-NN & \cite{asish_classification_2024, kuvar2023keystroke, MMSART2022eeg, singha2021gaze, bixler2021crossed, chang2021efficient, dias_da_silva_wandering_2020, faber2018automated, bixler2016automatic, bixler2014toward}\\
Proba- bilistic & Bayesian Networks, Hidden Markov Models & \cite{MMSART2022eeg, lee_when_2021, singha2021gaze, bixler2021crossed, chang2021efficient, dias_da_silva_wandering_2020, faber2018automated, hutt2017out, stewart2017, hutt2016eyes, bixler2016automatic, mills2016automatic, bixler2015, bixler2015automatic, bixler2014toward}\\
\rowcolor[gray]{0.9} 
Neural networks & CNN, LSTM, MLPs, DNN & \cite{pain_msstnet_2025, buhler_temporal_2025, asish_classification_2024, long_multimodal_2024, withammer-ekerhovd_classifying_2024, buhler_detecting_2024, buhler2024lab, jin2023decoding, china2023eeg, lee_predicting_2022, asish_detecting_2022, khan2022execute, zhu2022topographynet, moy_classification_2021, colorado2021, liu2021fnirsbased, singha2021gaze, guo2019deep, kane_combined_2017, hutt2016eyes}\\
\bottomrule
\end{tabular}
\label{tab:review_ml}
\end{table}

\paragraph{\textbf{Tasks}} 
With the widespread adoption of online learning environments, such as Massive Open Online Courses (MOOCs) that deliver high-quality educational content to a global audience~\citep{dabbagh2016massive, ferreira2016massive, fini2009technological, bodonhelyi2025passive}, supporting students in maintaining focus has become a pressing need. 
The reviewed studies predominantly focus on detecting mind wandering within common educational tasks, including reading~\citep{bixler2016automatic, mills2016automatic, bixler2015automatic, singha2021gaze, bixler2021crossed, UZZAMAN20111882, mills_eye-mind_2021, jaiyeola_one_2025, moy_classification_2021, jin2019predicting, buhler2024lab, colorado2021, bixler2015, brishtel2020}, lecture watching~\citep{khan2022feature, dhindsa2019individualized, china2023eeg, stewart2017, buhler_detecting_2024, kane_combined_2017}, writing or editing~\citep{rahnuma2024gazebased}, coding~\citep{rahnuma2024gazebased}, preparing presentations~\citep{rahnuma2024gazebased}, browsing online resources~\citep{rahnuma2024gazebased}, and engaging in computer-mediated conversations~\citep{kuvar2023keystroke}. More advanced tasks, such as those involving intelligent tutoring systems~\citep{hutt2016eyes, khan2022execute, asish_detecting_2022, vortmann_eeg-based_2019, colorado2021, asish_classification_2024, vortmann_eeg-based_2019} are also explored, reflecting efforts to mirror real-world learning environments as closely as possible.
While standardized cognitive tasks like the Sustained Attention to Response Task (SART)~\citep{MMSART2022eeg, lee_when_2021, withammer-ekerhovd_classifying_2024, jin2023decoding, liu2021fnirsbased, jin2020distinguishing, chang2021efficient, jin2019predicting}, visual search/monitoring~\citep{withammer-ekerhovd_classifying_2024, chiossi_designing_2025, long_multimodal_2024}, N-Back~\citep{long_multimodal_2024, chiossi_designing_2025}, breath counting~\citep{pain_msstnet_2025}, and auditory target detection~\citep{dong2021detection} may appear less directly tied to educational practice, they nonetheless tap into core cognitive processes such as sustained attention, working memory, and perceptual discrimination that underpin effective learning. Non-instructional tasks, including operation span \citep{dias_da_silva_wandering_2020}, short games \citep{moy_classification_2021}, and motor tasks \citep{aliakbaryhosseinabadi_classification_2017}, possess educational relevance, which is derived from their engagement and measurement of fundamental cognitive abilities linked to educational outcomes, such as cognitive control, problem-solving, and fine motor coordination.

\begin{tcolorbox}[mynote,title={\textbf{Current Modalities, Algorithms, and Learning Impact}}]
EEG and eye tracking remain dominant modalities in educational mind wandering research, with recent trends showing increased use of neural networks and real-world learning tasks. Despite promising developments, a systematic link between detection performance and learning outcomes remains underexplored, underscoring the need for domain-specific benchmarks and learner-centered applications.
\end{tcolorbox}

\subsubsection{Recent Advances}
This section reviews 17 works published subsequent to the March 2023 review by \citet{kuvar2023detecting}. While these studies follow established trends in utilizing EEG~\citep{riby_elevated_2025, pain_msstnet_2025, chiossi_designing_2025} and eye-tracking~\citep{jaiyeola_one_2025, buhler_temporal_2025, li_catching_2024} they also introduce key advancements. 

\paragraph{\textbf{Environments}}
There has been a growing diversification of environments for mind wandering detection. While virtual reality (VR) \citep{asish_detecting_2022} and augmented reality (AR)~\citep{vortmann_eeg-based_2019} were only sparsely explored before 2023, recent work~\citep{chiossi_designing_2025, asish_classification_2024, long_multimodal_2024} increasingly adopts VR-based learning tasks, often combining multimodal signals such as EEG and eye-tracking~\citep{long_multimodal_2024, asish_classification_2024}. In contrast to earlier studies relying on artificial external distractions~\citep{asish_detecting_2022}, newer approaches focus on detecting internally driven, more naturalistic mind wandering and, in some cases, enable real-time adaptive responses to learners’ attentional states~\citep{chiossi_designing_2025}, reflecting a shift toward more ecologically valid and personalized learning environments.

\paragraph{\textbf{Learner Groups}}
Earlier works highlighted the potential influence of learner characteristics such as attention deficit hyperactivity disorder (ADHD)~\citep{khan2022execute}, anxiety~\citep{khan2022feature}, and depression~\citep{khan2022feature} on mind wandering, suggesting it as a direction for future research. A recent study~\citep{jaiyeola_one_2025} addresses this gap by examining mind wandering detection across diverse learner groups, including neurodivergent profiles, showing that personalized models improve accuracy and behavioral insight—though this remains an open research challenge. 

\paragraph{\textbf{Detection Algorithms}}
One notable development in automatic mind wandering detection is the use of a multi-stream spatio-temporal deep learning architecture, MSSTNet~\citep{pain_msstnet_2025}, which has proven to be effective on EEG data. By integrating CNN-LSTM modules, this approach captures fine-grained spatio-temporal features across different frequency bands.
Another promising direction~\citep{rahnuma2024gazebased} integrates thought sampling and eye-tracking with optimized particle swarm algorithms to predict naturalistic thought dimensions, such as task-relatedness and internal–external orientation, from eye movement features.
These innovations mark a shift toward sophisticated and context-sensitive frameworks, yielding more accurate mind wandering assessments than traditional methods.

\paragraph{\textbf{Cross-Dataset Evaluation}}
Cross-dataset evaluation, which was previously hindered by challenges in an EEG-based study~\citep{jin2023decoding}, has recently shown improved performance in video-based contexts~\citep{buhler2024lab}, suggesting increased generalizability of models across diverse learning scenarios. This recent study showed that models trained on facial video data from lab-based reading tasks can transfer to naturalistic lecture viewing, achieving above-chance cross-dataset performance with bi-LSTM architectures~\citep{hochreiter1997long, baldi1999exploiting, schuster1997bidirectional}, and identifying facial regions such as the eyes and mouth as informative features.

\paragraph{\textbf{Multimodal Datasets}}
In the past two years, the number of available multimodal datasets for mind wandering detection has approximately doubled, reflecting a growing interest in capturing the complexity of attentional states through multiple synchronized signals. Earlier datasets primarily combined EEG~\citep{brishtel2020, MMSART2022eeg} or eye tracking~\citep{bixler2015, MMSART2022eeg} data with physiological signals such as EDA~\citep{brishtel2020}, skin conductance~\citep{bixler2015}, temperature~\citep{bixler2015}, photoplethysmography~\citep{MMSART2022eeg} or galvanic skin response signals~\citep{MMSART2022eeg}. In contrast, recent datasets combine modalities such as EEG with eye tracking~\citep{asish_classification_2024, long_multimodal_2024}, EEG with blink dynamics~\citep{riby_elevated_2025}, and eye tracking–based facial video with EDA and heart rate~\citep{buhler_detecting_2024}. Notably, \citet{buhler_detecting_2024} provide a rich multimodal dataset showing that combined modalities outperform single-sensor setups and enabling separate classification of aware and unaware mind wandering.

\paragraph{\textbf{Future directions}}
Despite recent progress, several gaps remain. Real-world integration of mind wandering detection into educational tools is still limited, constraining practical impact. Future work should advance technical robustness through improved feature fusion, domain adaptation, and self-supervised learning, alongside the expansion of multimodal datasets collected in ecologically valid and diverse learning contexts. Further research is needed to better capture the heterogeneity of mind wandering, such as aware versus unaware states and task-related versus task-unrelated thoughts - both in neurotypical and neurodivergent learners, where these distinctions remain insufficiently understood. In addition, the field lacks a unified framework for comparing detection approaches across modalities and settings. Finally, the growing use of sensitive behavioral and physiological data underscores the need for strong privacy-preserving and ethical design principles to support trustworthy deployment.

\begin{tcolorbox}[mynote,title={\textbf{Recent Progress and Emerging Frontiers}}]
Our analysis of recent research reveals significant advancements in environments, learner personalization, algorithm sophistication, and multimodal dataset development, yet highlights ongoing gaps in real-world applicability, cross-modal benchmarking, and privacy-aware deployment.
\end{tcolorbox}

\subsubsection{Discussion of the Systematic Review}
\paragraph{\textbf{Influence of mind wandering on learning outcomes}} Despite growing interest in mind wandering detection, the relationship between automatically detected mind wandering episodes and their impact on learning comprehension remains largely unexplored. While most studies, including those reviewed in this work~\citep{jaiyeola_one_2025, buhler_temporal_2025, mills_eye-mind_2021}, emphasize the detrimental effects of mind wandering on comprehension, it is plausible that not all off-task thoughts are equally harmful. More fine-grained distinctions like aware and unaware mind wandering \cite{buhler_detecting_2024} or intentional and unintentional mind wandering \cite{bosch2019automatic} remain largely underexplored. Such nuanced patterns of attentional fluctuation and their diverse effects on learning outcomes or the need for different types of interventions have yet to be systematically studied. 

\paragraph{\textbf{Applied algorithms}} While diverse detection algorithms—ranging from classical machine learning to advanced multimodal and deep learning architectures—have shown promise in identifying mind wandering, their effectiveness can vary substantially across datasets, modalities, and experimental contexts. 
Previous approaches have relied on centralized learning, which requires collecting sensitive physiological and behavioral data; this not only raises serious privacy concerns but also underscores the need to directly integrate privacy-preserving approaches to ensure compliance with regulations such as the AI Act~\citep{AIAct}.
Furthermore, evaluation protocols must move beyond methodologies that inadvertently bias results by including data from the same participants in both the training and testing sets \citep{long_multimodal_2024, moy_classification_2021, dhindsa2019individualized, china2023eeg, jin2019predicting}, as this practice can lead to overestimated performance metrics and misrepresent the model’s true generalizability. Only rigorous, user-independent evaluations can provide realistic estimates of model robustness. 

\section{Modular Benchmarking} 
In this section, we present a benchmarking study to identify effective mind wandering detection algorithms across sensing modalities. Due to the scarcity of datasets linking mind wandering to learning outcomes, we focus on detection performance as a foundation for future real-time educational interventions to support learning and comprehension.
To ensure fair and consistent comparisons, all models are assessed using the most commonly reported evaluation techniques from prior work~\citep{kuvar2023detecting}, including mind wandering $F_1$ score, above chance level $F_1$ score (\ref{app:ac_metric}), and Area Under the Curve (AUC). As previous studies frequently report multiple performance metrics to promote transparency, we follow this practice to allow nuanced performance assessment. Due to strong class imbalance, we use the mind wandering $F_1$ score and above-chance $F_1$ as primary metrics, as they better reflect performance than accuracy. These measures are especially important for real-world use, where both detecting mind wandering and limiting false alarms are essential for usability and trust.
To support ongoing research in this field, we developed and openly share a \textbf{{modular benchmarking framework}} (available on \href{https://gitlab.lrz.de/hctl/mw-benchmarking/}{GitLab}) that enables flexible application to both public and private datasets.

\subsection{Methodology}
To evaluate the performance of mind wandering detection algorithms across diverse data sources, we collected and curated a set of datasets identified through our systematic literature review. We used publicly available datasets and also contacted the authors to request access and successfully gathered seven EEG datasets~\citep{china2023eeg, dong2021detection, jin2020distinguishing, dhindsa2019individualized, jin2019predicting}, two facial video datasets~\citep{colorado2021, lee_predicting_2022}, and one multimodal dataset~\citep{buhler_detecting_2024} comprising facial video, eye tracking, EDA, and blood volume pressure. We collected a new eye-tracking dataset (available on \href{https://zenodo.org/records/14097753?preview=1&token=eyJhbGciOiJIUzUxMiJ9.eyJpZCI6IjAyM2QxNWY1LWJlY2ItNDQ2ZC04YjNhLWUwYjdkZGU0MzU4ZCIsImRhdGEiOnt9LCJyYW5kb20iOiJkNTkyZmI0MzMxMWRiODVjNzJkNzhkNGYyOWEzMjMwYiJ9.MRLZkGqYIwafNVI-D9z1de1qY4o-iQ1fPJFK5NXgImrL095j3cWAoMhwjJHlbD3ZdeVWrmDuPH_ItHQY-wGMrQ}{Zenodo}) (\ref{app:eye_dataset}) to enable more comprehensive and standardized comparisons across detection models.

\begin{figure}[t]
    \centering
    \includegraphics[width=14cm]{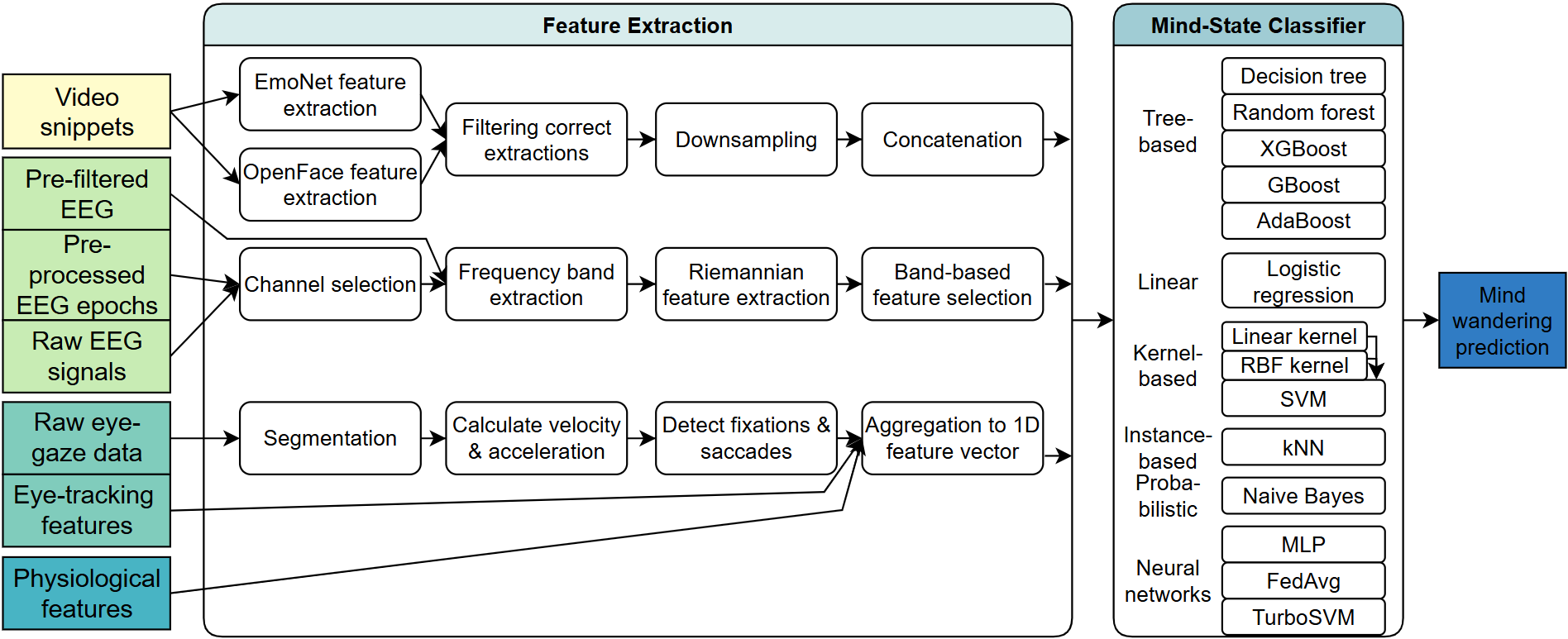}
    \caption{Data preprocessing and model architecture. }
    \label{fig:methodology}
\end{figure}

To ensure consistent and meaningful representations across diverse data types, we implemented a tailored yet generalizable feature extraction pipeline for each input modality. As shown in \autoref{fig:methodology}, different preprocessing strategies were applied depending on the input format, including video snippets, EEG signals (raw, preprocessed, or pre-filtered), eye-gaze recordings, and physiological data. For facial videos, features were extracted using OpenFace 2.2.0 \citep{baltrusaitis2018openface} for landmarks, action units, and gaze estimation, alongside EmoNet \citep{toisoul2021estimation} for emotion-related latent features. Depending on the dataset, we utilized either pre-clipped videos or manually extracted segments based on provided timestamps (10s long), adhering strictly to the original collectors' clipping protocols to ensure label consistency. These features were subsequently filtered, downsampled, and concatenated. The EEG data, which had been previously band-pass filtered (1–50 Hz), notch-filtered for line noise, and downsampled to 256 or 512 Hz, underwent a custom processing pipeline. Given the use of the probe-caught method across all datasets, signals were first aligned with probe triggers, retaining only the pre-probe epochs. Epoch durations were maintained according to the original study specifications to account for heterogeneous experimental settings. Subsequently, EEG data underwent a Riemannian geometry-based pipeline~\citep{china2023eeg}: selected channels were decomposed into frequency bands (delta, theta, alpha, beta), and features were extracted from covariance matrices in the Riemannian manifold. For eye-gaze data, temporal segments were analyzed to compute velocity and acceleration, detect fixations and saccades, and generate aggregated feature vectors. Physiological signals were similarly aggregated. This modular and scalable approach allowed modality-specific preprocessing while maintaining a unified structure across the multimodal dataset.

Our benchmark includes a diverse set of classification algorithms encompassing both traditional machine learning and neural methods (see \ref{app:algorithms} for details); the model selection was guided by insights from our systematic review (\autoref{tab:review_ml}), ensuring that at least one representative model from each major algorithmic category was included. To reflect their continued popularity in the literature, we implemented multiple tree-based methods, such as decision trees (DT)~\citep{breiman_classification_2017}, random forests (RF)~\citep{breiman2001random}, XGBoost~\citep{chen2016xgboost}, Gradient boosting (GBoost)~\citep{friedman2001greedy} and AdaBoost~\citep{hastie_multi-class_nodate}, and SVMs \citep{cortes1995support} with both linear and radial basis function (RBF) kernels, while also incorporating several neural network architectures (multilayer perceptron (MLP)~\citep{rumelhart_learning_1986}, models with stochastic gradient descent (SGD)~\citep{robbins1951stochastic} optimization and ReLU activation~\citep{glorot2011deep}) to capture recent trends favoring deep learning approaches in mind wandering detection. For federated learning experiments \citep{bodonhelyi2026safeguarding}, we utilized an MLP architecture as the base model and employed both FedAvg~\citep{McMahan2017a} and TurboSVM~\citep{Wang2024} aggregation strategies to simulate privacy-preserving, distributed learning across participants. We also included logistic regression (LR)~\citep{cox1958regression}, naïve Bayes (NB)~\citep{zhang_optimality_nodate}, and K-NN~\citep{cover_nearest_1967}. We implemented two supervised feature selection methods for the traditional classifiers: ANOVA~\citep{anova1989}, which uses the $F$-statistic to assess feature relevance, and mutual information~\citep{battiti_using_1994}, which captures non-linear dependencies. Hyperparameters were optimized for both the classifiers and selection methods. Model performance was evaluated using standard classification metrics: above chance level $F_1$ score (AC), where the sign of the values represents whether it is above or below the chance level, $F_1$ score, accuracy (Acc.), precision (Prec.), recall (Rec.) and AUC.

\begin{tcolorbox}[mynote,title={\textbf{Modular Benchmarking Framework}}]
We release a flexible, ready-to-use benchmarking framework that streamlines data preprocessing, model training, and evaluation across four sensing modalities, which is also easily adaptable to new datasets, features, and experimental needs.
\end{tcolorbox}

\subsection{Datasets}
We conducted our experiments across 14 datasets, encompassing a diverse range of modalities and settings; for an overview of each dataset's characteristics, see \autoref{tab:mw_summary}. These datasets represent all publicly available or author-accessible educational mind wandering datasets identified through our systematic review at the time of this study. While the included datasets span different educational tasks with varying levels of complexity (e.g., reading, lecture watching, interactive activities), all were selected for their relevance to learning contexts. To account for this heterogeneity, we explicitly report the underlying task alongside the results for each modality, enabling task-aware interpretation of performance differences. Although mind wandering is increasingly viewed as a heterogeneous phenomenon \citep{buhler_detecting_2024}, most publicly available datasets provide only binary labels. Therefore, to ensure comparability across datasets, our benchmark focuses on binary detection, while recognizing the need for richer, multi-dimensional annotations in future work.

\begin{table}[t]
\centering
\footnotesize
\caption{Overview of the used datasets after data preprocessing. Abbreviations: MW - number of positive samples, GT - ground truth collection method, Env. - Data collection environment.  }
\begin{tabular}{p{25mm} p{13mm} p{14mm} p{12mm} p{8mm} p{6mm} p{8mm} p{8mm} p{10mm}}
\toprule
\textbf{Dataset} & \textbf{Modality} & \textbf{Task} & \textbf{Samples} & \textbf{MW} & \textbf{Users} & \textbf{GT} & \textbf{Env.} & \textbf{MW ratio}\\
\midrule
China       & EEG      & watch  & 12500 & 5941 & 14 & self  & lab & .475      \\
California  & EEG      & listen            & 301   & 158  & 13 & probe & lab & .525      \\
Neth2-SART  & EEG      & SART               & 291   & 126  & 12 & probe & lab & .433      \\
Neth2-VS   & EEG      & VS      & 869   & 374  & 19 & probe & lab & .430      \\
Neth1-SART  & EEG      & SART               & 6888  & 3252 & 26 & probe & lab & .472     \\
Neth1-VS    & EEG      & VS      & 6240  & 2586 & 23 & probe & lab & .414      \\
Canada      & EEG      & watch     & 755   & 279  & 15 & probe & class & .370 \\
\midrule
Colorado    & video & read         & 2519  & 944  & 94 & self  & lab & .375      \\
Korea       & video & watch  & 797  & 194  & 10  & probe & home & .243     \\
\midrule
Ours    & eye gaze  & watch       & 1344  & 628  & 43  & self  & lab & .467      \\
New Hampshire & eye gaze & read/listen & 1410 &  435 & 47 & probe & home &  .308\\
USA & eye gaze & read & 777 & 292 & 111 & self & home & .376 \\
Online & eye gaze & watch & 475 & 194 & 95 & - & home & .408  \\
\midrule
Germany      & multi & watch   & 791   & 320  & 66  & probe & lab & .405      \\
\bottomrule
\end{tabular}
\label{tab:mw_summary}
\end{table}

\paragraph{EEG Datasest} 
The \textbf{China dataset} \citep{china2023eeg} was recorded during video-based learning under two conditions (focused learning vs. future planning) using an 8-channel EEG, with Riemannian spatial covariance features and self-caught keypress labels enabling accurate within-subject RBF-SVM classification. The \textbf{Netherlands-1} \citep{jin2019predicting} and \textbf{Netherlands-2} \citep{jin2020distinguishing} datasets combined Sustained Attention to Response Task (SART) and visual search (VS) - which allowed us to separate the datasets based on these tasks - to detect mind wandering through probe-based labeling and alpha-band feature extraction; however, generalization remained limited. The \textbf{Canada dataset} \citep{dhindsa2019individualized}, recorded during real university lectures with a 16-channel setup, showed strong within-subject but weak cross-subject performance using non-linear SVMs. Finally, the \textbf{California dataset} \citep{dong2021detection}, collected during an auditory detection task with 64-channel EEG, provided event-related potential features for above-chance cross-subject AUC, though only extracted features were available for analysis.

\paragraph{Video Datasets} 
The \textbf{Colorado dataset}\citep{colorado2021} was collected in a laboratory during a computerized reading task, using webcam-recorded facial videos and self-caught mind wandering labels. Ten-second video segments preceding mind wandering reports were labeled accordingly, while non-mind wandering clips were sampled earlier to maintain a 30\% mind wandering ratio. The best performance was achieved with an SVM classifier, yielding an $F_1$ score of 0.48. The \textbf{Korea dataset}\citep{lee_predicting_2022} was recorded during an online lecture, also using webcam-based facial videos, with multiple pre-probe window lengths (5, 10, 20 seconds) tested. The top-performing configuration employed XGBoost with a 20-second window, achieving an $F_1$ score of 0.33, though the dataset’s small size and class imbalance (20\% mind wandering) limited generalizability.

\paragraph{Eye-Tracking Datasets} 
\textbf{Our eye-tracking dataset} (available on \href{https://zenodo.org/records/14097753?preview=1&token=eyJhbGciOiJIUzUxMiJ9.eyJpZCI6IjAyM2QxNWY1LWJlY2ItNDQ2ZC04YjNhLWUwYjdkZGU0MzU4ZCIsImRhdGEiOnt9LCJyYW5kb20iOiJkNTkyZmI0MzMxMWRiODVjNzJkNzhkNGYyOWEzMjMwYiJ9.MRLZkGqYIwafNVI-D9z1de1qY4o-iQ1fPJFK5NXgImrL095j3cWAoMhwjJHlbD3ZdeVWrmDuPH_ItHQY-wGMrQ}{Zenodo}; more information is shared in \ref{app:eye_dataset}) was collected to investigate mind wandering during educational video viewing. Participants
($N = 56$, $N_{male}=14$, $N_{female}=42$, $M_{age}=26.23, SD_{age}=5.50$) with normal or corrected vision watched three instructional videos on derivatives while their gaze was recorded at 1000 Hz using an EyeLink 1000 Plus system. Mind wandering was self-caught via key presses, and a retention test one week later enabled linking attentional lapses to learning outcomes. The \textbf{New Hampshire dataset} \citep{hutt2024webcam} was recorded online from 173 participants using webcam-based eye tracking (WebGazer~\citep{papoutsaki2016webgazer}) during a 40-minute narrative anticipation task with probe-caught labeling, achieving an $F_1$ score of 25\% with XGBoost. The \textbf{USA dataset} \citep{jaiyeola_one_2025} also used WebGazer for reading tasks completed at home. Models trained on neurotypical learners with gaze and NLP-based features achieved an AUROC of 52\%. Finally, the \textbf{Online dataset} ~\citep{steadman2025difficulty} involved participants watching an educational video while their gaze behavior was analyzed over 20-second pre-probe windows, focusing on task-unrelated thoughts, disengagement, valence, and boredom.

\paragraph{Multimodal Datasets}
\textbf{Germany multimodal dataset} \citep{buhler_detecting_2024} was collected in a controlled lab study at a German university, where participants watched a one-hour educational lecture while facial videos, eye tracking, and physiological data were recorded. Mind wandering (40\%) was labeled via 15 probe-caught questions per participant. From 30-second pre-probe windows, 804 facial, 155 gaze, and 108 physiological features were extracted. The best unimodal AUC-PRs were 52.6\% (facial, XGBoost), 50.3\% (gaze, SVM), and 45.7\% (physiological, MLP). Multimodal fusion achieved the highest performance (63.7\% AUC-PR, $F_1$=58.0\%), highlighting the advantage of integrating modalities.

\subsection{Experimental Setup}

\begin{table}[t]
    \centering
    \scriptsize
    \caption{Overview of tuned hyperparameters and their search ranges.}
    \begin{tabular}{c c c}
    \toprule
    \textbf{Hyperparameter} & \textbf{ML Algorithm} & \textbf{Range} \\
    \midrule
    \multirow{2}{*}{Regularization} & SVM & [0.1, 1, 10, 100] \\
     & LR & [0.1, 1, 10] \\
     \rowcolor[gray]{0.9} 
    Kernel coefficient & SVM (RBF) & ['scale', 'auto', 0.001, 0.01, 0.1] \\
    Class weights & SVM, LR, RF, DT & [None, 'balanced'] \\
    \rowcolor[gray]{0.9} 
    Penalty type & LR & ['l1', 'l2'] \\
    Variance smoothing & Naive Bayes & [1e-9, 1e-8, 1e-7, 1e-6] \\
    \rowcolor[gray]{0.9} 
    Num. of neighbors & kNN & [3, 5, 7, 9] \\
    Distance metric & kNN & ['euclidean', 'manhattan'] \\
    \rowcolor[gray]{0.9} 
    Neighbors' weight & kNN & ['uniform', 'distance'] \\
    \multirow{3}{*}{Num. of trees} & RF & [100, 200, 300] \\ 
    & XGBoost, GBoost & [100, 200] \\
    & AdaBoost & [50, 100, 200] \\
    \rowcolor[gray]{0.9} 
     & RF, DT & [None, 10, 20, 30] \\
    \rowcolor[gray]{0.9} 
    \multirow{-2}{*}{Max. tree depth} & XGBoost, GBoost & [3, 6, 9] \\
    Min. samples to split & RF, DT & [2, 5, 10] \\
    \rowcolor[gray]{0.9} 
    Min. samples/leaf & DT & [1, 2, 4] \\
    \multirow{3}{*}{Learning rate} & XGBoost, GBoost, AdaBoost & [0.01, 0.1, 0.3]\\
     & MLP & [1e-3, 1e-4, 1e-5] \\
    & FedAvg & [1e-2, 1e-3, 1e-4]  \\
    \rowcolor[gray]{0.9} 
    Subsampling rate & XGBoost, GBoost & [0.8, 1.0] \\
    Column subsample & XGBoost & [0.8, 1.0] \\
    \rowcolor[gray]{0.9} 
    Batch size & MLP, FedAvg, TurboSVM & [4, 8, 16] \\      
    Client epochs &  FedAvg, TurboSVM & [5, 15]  \\
    \rowcolor[gray]{0.9}
    Clients/round & FedAvg, TurboSVM & [0.5, 1.0] \\
    Logits learning rate & TurboSVM & [1e-3, 1e-4, 1e-5] \\
    \bottomrule
    \end{tabular}
    \label{tab:hparam_ranges}
\end{table}

For each model and dataset, we first determined hyperparameters before running the final evaluations. For traditional machine learning classifiers, we performed 5-fold cross-validation with $seed=0$ on the combined training and validation sets (\autoref{tab:hparam_ranges}), keeping the test set strictly unseen during tuning. This was computationally feasible due to the relatively fast training times. In contrast, for MLP and federated learning approaches, which required substantially longer training, hyperparameters were selected using a predefined grid search over three seeds, and Bayesian optimization was applied for federated models. Using the chosen configurations, experiments were then repeated five times with different random seeds ${0,1,2,3,4}$ on a fixed 80\%–10\%–10\% person-independent train–validation–test split, and we report the mean and standard deviation across runs. We used early stopping to prevent overfitting, with a patience of 10 epochs.
The number of hidden layers was adapted to feature dimensionality: one for the California dataset (4 ERP features), New Hampshire, USA and Online eye-gaze datasets (7 gaze features), two for band-specific EEG (36 features), and three for concatenated bands, eye-tracking, physiological, or multimodal inputs (144–546 features). We used ReLU activations, dropout (0.1), and batch normalization throughout.
Both FedAvg and TurboSVM employed the same base MLP; for TurboSVM, we reused FedAvg parameters, tuning only the number of global epochs and server learning rate.

\subsection{Results}
To provide a comprehensive overview of model performance across modalities and datasets, we report the best-performing classifiers in three categories: the top traditional machine learning model, the MLP model, and the leading federated learning model. Performance is summarized for each input modality (EEG, facial video, eye tracking, and physiological signals), enabling a modality-specific comparison of model effectiveness. Because the benchmark spans multiple educational tasks, results are expected to vary across datasets. Accordingly, we report performance in a task- and modality-aware manner rather than aggregating results across heterogeneous contexts.

\begin{table}[t]
    \centering
    \scriptsize
    \caption{Summary of the best-performing models for each EEG dataset based on the mind wandering ($F_1$) score, comparing traditional machine learning models, MLPs, and federated learning approaches (FedAvg and TurboSVM). Results are reported for individual EEG frequency bands as well as for all bands combined, with additional metrics such as precision and recall included for completeness.}
    \begin{tabular}{p{7mm} p{14mm} p{7mm} p{13mm} p{6mm} p{14mm} p{15mm} p{14mm} p{13mm}}
    \toprule
         \raisebox{-0.5\height}{\textbf{Dataset}} & \raisebox{-0.5\height}{\textbf{Best model}} & \textbf{Best band(s)} & \raisebox{-0.5\height}{$\mathbf{F_1}$ \textbf{[\%]}} & \raisebox{-0.5\height}{\textbf{AC [\%]}} & \raisebox{-0.5\height}{\textbf{Prec. [\%]}} &\raisebox{-0.5\height}{\textbf{Rec.[\%]}} &  \raisebox{-0.5\height}{\textbf{AUC [\%]}} & \raisebox{-0.5\height}{\textbf{Acc. [\%]}}\\
    \midrule
         \multirow{3}{*}{China} & Tree & theta & 56.1$\pm$0.1 & 15.5 & 48.4$\pm$0.0 & 66.7$\pm$0.3 & 51.4$\pm$0.1 & 49.8$\pm$0.0
         \\
          & MLP & all & 58.0$\pm$6.3 & 19.1 & 51.3$\pm$3.9 & 71.7$\pm$20.6 & 52.8$\pm$5.3 & 52.4$\pm$3.5 \\
         & \textbf{TurboSVM} & all & \textbf{63.5$\pm$1.7} & \textbf{29.7} & 51.0$\pm$3.1 & 85.6$\pm$8.2 & 59.1$\pm$4.6 & 52.9$\pm$4.6 \\
         \midrule
         \multirow{3}{*}{\thead{Cali- \\ fornia}} & SVM-lin. & all & 58.3$\pm$0.0 & 8.7 & 60.9$\pm$0.0 & 56.0$\pm$0.0 & 55.3$\pm$7.3 & 56.5$\pm$0.0 \\
          & MLP & all & 55.1$\pm$13.5 & 1.6 & 58.1$\pm$5.3 & 58.4$\pm$23.5 & 51.2$\pm$11.3 & 53.5$\pm$5.1 \\
          & \textbf{TurboSVM} & all & \textbf{59.4$\pm$10.3} & \textbf{11.1} & 55.2$\pm$6.2 & 68.0$\pm$21.9 & 50.4$\pm$6.4 & 52.2$\pm$9.2 \\
         \midrule
         \multirow{3}{*}{\thead{Neth2 \\ SART}} & \textbf{SVM-rbf} & all & \textbf{85.7$\pm$0.0} & \textbf{75.3} & 84.0$\pm$0.0 & 87.5$\pm$0.0 & 89.1$\pm$0.0 & 87.7$\pm$0.0 \\
          & MLP & theta & 54.3$\pm$11.1 & 21.1 & 49.4$\pm$9.9 & 64.2$\pm$19.1 & 61.0$\pm$11.9 & 56.6$\pm$11.0 \\
          & TurboSVM & alpha & 68.7$\pm$3.2 & 45.9 & 62.2$\pm$2.6 & 77.5$\pm$8.6 & 74.4$\pm$5.9 & 71.0$\pm$2.0 \\
         \midrule
         \multirow{3}{*}{\thead{Neth2 \\ VS}} & \textbf{LR} & alpha & \textbf{61.2$\pm$0.0} & \textbf{31.0} & 44.6$\pm$0.0 & 97.6$\pm$0.0 & 50.7$\pm$0.0 & 45.8$\pm$0.0 \\
          & MLP & all & 51.9$\pm$10.2 & 14.5 & 43.6$\pm$4.2 & 67.6$\pm$22.0 & 47.5$\pm$5.2 & 48.5$\pm$4.1 \\
          & FedAvg & all & 55.9$\pm$4.2 & 21.6 & 42.5$\pm$2.3 & 81.9$\pm$10.4 & 49.3$\pm$5.0 & 44.0$\pm$3.0 \\
         \midrule
         \multirow{3}{*}{\thead{Neth1 \\ SART}} & \textbf{NB} & delta & \textbf{60.6$\pm$0.0} & \textbf{25.9} & 57.0$\pm$0.0 & 64.7$\pm$0.0 & 63.3$\pm$0.0 & 60.6$\pm$0.0 \\
          & MLP & all & 58.8$\pm$5.1 & 22.5 & 48.6$\pm$2.6 & 78.5$\pm$17.8 & 51.7$\pm$3.1 & 50.1$\pm$3.1 \\
          & FedAvg & all & 58.9$\pm$3.8 & 22.7 & 48.4$\pm$2.2 & 77.9$\pm$15.0 & 50.7$\pm$2.7 & 50.1$\pm$2.9 \\
         \midrule
         \multirow{3}{*}{\thead{Neth1 \\ VS}} & LR & all & 55.8$\pm$0.0 & 23.7 & 43.0$\pm$0.0 & 79.2$\pm$0.0 & 53.8$\pm$0.0 & 47.1$\pm$0.0 \\
          & \textbf{MLP} & all & \textbf{56.6$\pm$2.6} & \textbf{25.0} & 43.4$\pm$1.1 & 82.0$\pm$8.9 & 52.5$\pm$2.4 & 47.3$\pm$2.2 \\
          & FedAvg & all & 54.2$\pm$3.7 & 20.9 & 43.0$\pm$1.4 & 73.9$\pm$9.7 & 52.4$\pm$2.7 & 47.9$\pm$1.7 \\
         \midrule
         \multirow{3}{*}{Canada} & \textbf{LR} & all & \textbf{53.7$\pm$0.0} & \textbf{28.1} & 40.7$\pm$0.0 & 78.6$\pm$0.0 & 54.6$\pm$0.0 & 51.7$\pm$0.0 \\
          & MLP & theta & 46.2$\pm$5.1 & 16.5 & 36.7$\pm$1.2 & 64.8$\pm$17.0 & 53.3$\pm$3.3 & 48.0$\pm$3.7 \\
          & TurboSVM & beta & 50.7$\pm$3.5 & 23.5 & 42.9$\pm$3.0 & 64.8$\pm$14.9 & 58.4$\pm$2.8 & 55.9$\pm$5.3 \\
    \bottomrule
    \end{tabular}
    \label{tab:eeg_res}
\end{table}

\paragraph{\textbf{EEG Results}} A comparison of our results (\autoref{tab:eeg_res}) with previous studies reveals important methodological differences that limit the direct comparability of performance metrics. In the China dataset, user-dependent evaluation was employed, resulting in higher reported performance, specifically, an SVM with RBF kernel trained on the concatenation of all frequency band features achieved an AUC of 87.6$\pm$7.0\%. In contrast, our user-independent approach yielded the best AUC of 59.1$\pm$4.6\%, underscoring the known inflation of performance in user-dependent settings. For the California dataset, a user-independent evaluation with leave-one-subject-out cross-validation was applied. After addressing class imbalance through oversampling in the training set, their SVM with RBF kernel achieved 59.1$\pm$7.0\% accuracy and an AUC of 61.3$\pm$8.5\%, aligning well with our results. In the Neth2 dataset, models were trained exclusively on the VS task and tested on SART data, a domain transfer setting that renders our user-independent evaluation not directly comparable. The Neth1 dataset involved user-dependent training, with individual models fit per participant, class imbalance corrected through oversampling, and performance validated via leave-one-out cross-validation. Reported accuracies ranged from 50\% to 85\% ($M_{SART} = 64\%$, $M_{VS} = 69\%$) using SVMs, while our cross-participant setup yielded slightly lower results (60.6$\pm$0.0\% on SART, 47.3$\pm$2.2 \% on VS), as the evaluation protocols differ substantially. Finally, in the Canada dataset, both our work and the original study used inter-subject classification with leave-one-subject-out cross-validation. While their non-linear SVM failed to exceed chance-level performance, our logistic regression classifier achieved an $F_1$ score 28.1\% above chance, indicating meaningful predictive capability in this challenging setting.

\begin{tcolorbox}[mynote,title={\textbf{EEG Results - Key Takeaway}}]
Across datasets, the best-performing approaches were mostly traditional machine learning models, often outperforming more complex architectures. Task-specific patterns emerged: lecture and video watching tasks favored traditional and federated approaches, listening tasks were best addressed by federated models, SART tasks benefited from traditional methods, and VS tasks showed particularly strong performance with logistic regression and MLPs. EEG frequency band analysis revealed that models trained on all bands jointly tended to yield the highest performance, although certain datasets showed strong results with theta or alpha bands alone.
\end{tcolorbox}

\paragraph{\textbf{Video Results}} In the video modality (\autoref{tab:video_res}), the original publication of the Colorado dataset reported a deep learning approach achieving an $F_1$ score of 42.1\% using features derived from OpenFace. Our logistic regression model, trained under a comparable setup (although with different feature set), reached increased performance with an $F_1$ score of 58.7 $\pm$ 0.0\%. In the same domain, \citet{buhler2024lab} reported $F_1$ scores of 45.9\% with explicit OpenFace-derived features (such as gaze and action units) and 45.3\% with latent affective representations using a bi-LSTM, further supporting the consistency of these results. For the Korea dataset, a deep learning model using eye and head movement, gaze, and emotion indicators reached 31 $\pm$ 8\% $F_1$ on the positive class. Under similar conditions but with different input features, our  SVM model with RBF kernel yielded 42.6$\pm$0.0\% $F_1$ score, while the MLP model achieved 24.1 $\pm$ 19.7\% $F_1$ score. In a cross-dataset evaluation, \citet{buhler2024lab} reported $F_1$ scores of 25.1\% for explicit features and 35.2\% for latent ones on the same dataset. Lastly, for the multimodal German dataset, the original authors obtained an $F_1$ of 49.3\% with XGBoost on video-based features, while our SVM model with linear kernel function achieved a higher score of 61.2 $\pm$ 0.0\%. Our findings indicate that the current neural network performance could be enhanced by refining the feature selection process to prioritize high-value indicators like gaze and head pose, while simultaneously adopting complex architectures like bi-LSTMs to model essential temporal dynamics.

\begin{tcolorbox}[mynote,title={\textbf{Video Results - Key Takeaway}}]
While federated learning offers strong privacy protection, traditional machine learning models also demonstrated strong performance and, in reading and video-watching tasks, consistently outperformed both deep and federated approaches, highlighting their suitability for real-time detection.
\end{tcolorbox}

\begin{table}[t]
    \centering
    \scriptsize
    \caption{Summary of the best-performing models for each video dataset based on the mind wandering $F_1$ score, using the same feature extraction and selection pipeline across all datasets and comparing traditional machine learning models, MLPs, and federated learning approaches (FedAvg and TurboSVM). Additional metrics, including precision and recall, are reported for completeness.}
    \begin{tabular}{p{15mm} p{15mm} p{12mm} p{9mm} p{14mm} p{12mm} p{13mm} p{12mm}}
    \toprule
         \raisebox{-0.5\height}{\textbf{Dataset}} & \raisebox{-0.5\height}{\textbf{Best model}} & \raisebox{-0.5\height}{$\mathbf{F_1}$ \textbf{[\%]}} & \raisebox{-0.5\height}{\textbf{AC [\%]}} & \raisebox{-0.5\height}{\textbf{Prec. [\%]}} &\raisebox{-0.5\height}{\textbf{Rec.[\%]}} &  \raisebox{-0.5\height}{\textbf{AUC [\%]}} & \raisebox{-0.5\height}{\textbf{Acc. [\%]}}\\
    \midrule
        \multirow{3}{*}{Colorado} & \textbf{LR} & \textbf{58.7$\pm$0.0} & \textbf{32.3} & 43.9$\pm$0.0 & 88.3$\pm$0.0 & 60.0$\pm$0.1 & 51.5$\pm$0.0 \\
         & MLP & 33.6$\pm$27.4 & -8.9 & 23.3$\pm$19.0 & 60.0$\pm$49.0 & 55.0$\pm$5.6 & 47.8$\pm$10.9 \\
         & FedAvg & 33.6$\pm$27.4 & -8.9 & 23.3$\pm$19.0 & 60.0$\pm$49.0 & 54.7$\pm$4.9 & 47.8$\pm$10.9 \\
         \midrule
        \multirow{3}{*}{Korea} & \textbf{SVM-rbf} & \textbf{42.6$\pm$0.0} & \textbf{23.3} & 35.4$\pm$0.0 & 53.5$\pm$0.0 & 55.5$\pm$0.0 & 63.7$\pm$0.0 \\
         & MLP & 24.1$\pm$19.7 & -1.4 & 15.1$\pm$12.3 & 60.0$\pm$49.0 & 52.9$\pm$11.2 & 45.0$\pm$24.4 \\
         & FedAvg & 24.1$\pm$19.7 & -1.4 & 15.1$\pm$12.3 & 60.0$\pm$49.0 & 56.8$\pm$9.5 & 45.0$\pm$24.4 \\
         \midrule
        \multirow{3}{*}{\thead{Germany \\ (video)}} & \textbf{SVM-lin.} & \textbf{61.2$\pm$0.0} & \textbf{32.9} & 52.0$\pm$0.0 & 74.3$\pm$0.0 & 33.2$\pm$0.0 & 60.2$\pm$0.0 \\
         & MLP & 35.7$\pm$29.2 & -11.2 & 25.7$\pm$21.0 & 58.9$\pm$48.1 & 52.4$\pm$6.6 & 49.4$\pm$7.1 \\
         & TurboSVM & 36.9$\pm$19.0 & -9.1 & 37.7$\pm$19.5 & 38.9$\pm$22.2 & 57.8$\pm$1.1 & 54.5$\pm$4.8 \\
    \bottomrule
    \end{tabular}
    \label{tab:video_res}
\end{table}

\begin{table}[t]
    \centering
    \scriptsize
    \caption{Summary of the best-performing models for each eye-tracking dataset based on the mind wandering $F_1$ score, comparing traditional machine learning models, MLPs, and federated learning approaches (FedAvg and TurboSVM). Additional metrics, including precision and recall, are reported for completeness.}
    \begin{tabular}{p{14mm} p{14mm} p{15mm} p{9mm} p{14mm} p{15mm} p{13mm} p{13mm}}
    \toprule
         \raisebox{-0.5\height}{\textbf{Dataset}} & \raisebox{-0.5\height}{\textbf{Best model}} & \raisebox{-0.5\height}{$\mathbf{F_1}$ \textbf{[\%]}} & \raisebox{-0.5\height}{\textbf{AC [\%]}} & \raisebox{-0.5\height}{\textbf{Prec. [\%]}} &\raisebox{-0.5\height}{\textbf{Rec.[\%]}} &  \raisebox{-0.5\height}{\textbf{AUC [\%]}} & \raisebox{-0.5\height}{\textbf{Acc. [\%]}}\\
        \midrule
        \multirow{3}{*}{Ours} & \textbf{LR} & \textbf{66.3$\pm$0.0} & \textbf{35.8} & 61.1$\pm$0.0 & 72.4$\pm$0.0 & 67.5$\pm$0.0 & 64.8$\pm$0.0 \\
         & MLP & 58.0$\pm$2.7 & 20.0 & 51.2$\pm$2.0 & 67.1$\pm$4.6 & 56.7$\pm$5.5 & 54.0$\pm$2.5 \\
         & TurboSVM & 61.1$\pm$4.2 & 25.9 & 57.9$\pm$2.1 & 65.0$\pm$7.2 & 62.3$\pm$4.2 & 60.8$\pm$2.6 \\
         \midrule
        \multirow{3}{*}{\thead{Germany \\ (gaze)}} & Tree & 51.3$\pm$1.3 & 15.8 & 53.0$\pm$1.4 & 49.7$\pm$1.4 & 61.4$\pm$1.7 & 60.2$\pm$1.1 \\
         & \textbf{MLP} & \textbf{54.6$\pm$12.5} & \textbf{21.5} & 49.9$\pm$5.5 & 67.4$\pm$26.6 & 61.8$\pm$5.6 & 57.1$\pm$6.4 \\
         & FedAvg & 52.2$\pm$13.5 & 17.3 & 47.5$\pm$7.0 & 64.0$\pm$26.6 & 60.9$\pm$5.9 & 55.4$\pm$5.8 \\
         \midrule
        \multirow{3}{*}{\thead{New \\ Hampshire}} & LR & 49.0$\pm$0.0 & 27.1 & 34.5$\pm$0.0 & 84.4$\pm$0.0 & 61.7$\pm$0.0 & 47.3$\pm$0.0 \\
         & \textbf{MLP} & \textbf{51.1$\pm$2.5} & \textbf{30.1} & 37.1$\pm$0.8 & 83.1$\pm$9.8 & 62.2$\pm$0.9 & 52.7$\pm$2.0 \\
         & FedAvg & 37.7$\pm$12.2 & 11.0 & 32.1$\pm$2.4 & 61.8$\pm$34.6 & 56.0$\pm$6.0 & 48.9$\pm$12.9 \\
         \midrule
        \multirow{3}{*}{USA} & \textbf{LR} & \textbf{52.8$\pm$0.0} & \textbf{25.2} & 37.3$\pm$0.0 & 90.3$\pm$0.0 & 50.6$\pm$0.0 & 40.5$\pm$0.0 \\
         & MLP & 47.8$\pm$3.8 & 17.3 & 38.4$\pm$3.5 & 63.9$\pm$7.2 & 51.5$\pm$5.3 & 48.6$\pm$5.3 \\
         & FedAvg & 39.0$\pm$11.0 & 3.3 & 40.4$\pm$10.2 & 39.4$\pm$15.0 & 51.9$\pm$5.5 & 56.2$\pm$6.9 \\
         \midrule
        \multirow{3}{*}{Online} & \textbf{SVM-lin.} & \textbf{54.0$\pm$0.0} & \textbf{20.7} & 40.5$\pm$0.0 & 81.0$\pm$0.0 & 47.9$\pm$0.0 & 42.0$\pm$0.0 \\
         & MLP & 52.1$\pm$4.4 & 17.4 & 41.4$\pm$3.7 & 70.5$\pm$7.0 & 47.7$\pm$4.5 & 45.6$\pm$5.9 \\
         & TurboSVM & 34.6$\pm$22.5 & -12.8 & 32.1$\pm$17.4 & 44.8$\pm$37.7 & 46.2$\pm$4.4 & 50.8$\pm$5.7 \\
    \bottomrule
    \end{tabular}
    \label{tab:et_res}
\end{table}

\paragraph{\textbf{Eye-Tracking Results}} Across the five eye-tracking datasets, MLPs yielded the best results in two cases, while logistic regression and SVMs outperformed other models in the remaining datasets. Importantly, the observed model performances did not appear to be tied to the specific task domains. To ensure comparability, data from neurodivergent participants were excluded in the New Hampshire and USA datasets. For the Germany dataset, under comparable conditions, we achieved an $F_{1}$ score of 54.6 $\pm$ 12.5\%, which is slightly higher than the 49.6\% reported by the original authors. For the other datasets, however, direct comparison with prior work is challenging due to limited publicly available performance details.

\begin{tcolorbox}[mynote,title={\textbf{Eye-Tracking Results - Key Takeaway}}]
The findings indicate that although performance varied across datasets, MLPs consistently yielded the strongest results, followed closely by logistic regression, with both approaches performing well across video-watching and reading tasks, suggesting that eye-tracking–based mind wandering detection is largely independent of these specific educational activities.
\end{tcolorbox}

\begin{table}[t]
    \centering
    \scriptsize
    \caption{Summary of the best-performing models for the physiological dataset based on the mind wandering $F_1$ score, comparing traditional machine learning models, MLPs, and federated learning approaches (FedAvg and TurboSVM). Additional metrics, including precision and recall, are reported for completeness.}
    \begin{tabular}{p{15mm} p{16mm} p{11mm} p{9mm} p{13mm} p{15mm} p{12mm} p{11mm}}
    \toprule
         \raisebox{-0.5\height}{\textbf{Dataset}} & \raisebox{-0.5\height}{\textbf{Best model}} & \raisebox{-0.5\height}{$\mathbf{F_1}$ \textbf{[\%]}} & \raisebox{-0.5\height}{\textbf{AC [\%]}} & \raisebox{-0.5\height}{\textbf{Prec. [\%]}} &\raisebox{-0.5\height}{\textbf{Rec.[\%]}} &  \raisebox{-0.5\height}{\textbf{AUC [\%]}} & \raisebox{-0.5\height}{\textbf{Acc. [\%]}}\\
    \midrule
        \multirow{3}{*}{\thead{Germany \\ (physio.)}} & \textbf{LR} & \textbf{55.6$\pm$0.0} & \textbf{23.2} & 54.1$\pm$0.0 & 57.1$\pm$0.0 & 65.4$\pm$0.0 & 61.4$\pm$0.0 \\
          & MLP & 40.3$\pm$17.9 & -3.2 & 42.5$\pm$9.1 & 42.3$\pm$20.8 & 51.8$\pm$8.9 & 54.9$\pm$2.2 \\
          & TurboSVM & 35.4$\pm$15.6 & -11.7 & 38.3$\pm$5.4 & 37.7$\pm$25.2 & 51.8$\pm$4.7 & 51.1$\pm$2.1 \\
    \bottomrule
    \end{tabular}
    \label{tab:phys_res}
\end{table}

\paragraph{\textbf{Physiological Results}} We evaluated the models on the only dataset in our benchmark (\autoref{tab:phys_res}) that included physiological signals, which was part of a larger multimodal dataset~\citep{buhler_detecting_2024}. The original authors reported their performance using an MLP with an $F_1$ score of 43.8\%, which is consistent with our findings, where the MLP achieved an $F_1$ score of 40.3 $\pm$ 17.9\% on the mind wandering class. However, in our experiments, the highest performance was obtained using logistic regression, reaching an $F_1$ score of 55.6 $\pm$ 0.0\%. This result outperformed the other tested traditional classifiers, including random forests, XGBoost, and SVMs, highlighting logistic regression’s effectiveness for this modality.

\begin{tcolorbox}[mynote,title={\textbf{Physiological Results - Key Takeaway}}]
For physiological data, traditional models, especially logistic regression, outperformed other approaches, but the scarcity of comparable datasets limits the generalizability of this finding.
\end{tcolorbox}

\begin{table}[t]
    \centering
    \scriptsize
    \caption{Summary of the best-performing models for the multimodal dataset using all modalities based on the mind wandering $F_1$ score, using the same feature extraction and selection pipeline across all datasets and comparing traditional machine learning models, MLPs, and federated learning approaches (FedAvg and TurboSVM). Additional metrics, including precision and recall, are reported for completeness.}
    \begin{tabular}{p{14mm} p{16mm} p{13mm} p{10mm} p{15mm} p{13mm} p{14mm} p{13mm}}
    \toprule
         \raisebox{-0.5\height}{\textbf{Dataset}} & \raisebox{-0.5\height}{\textbf{Best model}} & \raisebox{-0.5\height}{$\mathbf{F_1}$ \textbf{[\%]}} & \raisebox{-0.5\height}{\textbf{AC [\%]}} & \raisebox{-0.5\height}{\textbf{Prec. [\%]}} &\raisebox{-0.5\height}{\textbf{Rec.[\%]}} &  \raisebox{-0.5\height}{\textbf{AUC [\%]}} & \raisebox{-0.5\height}{\textbf{Acc. [\%]}}\\
    \midrule
        \multirow{3}{*}{\thead{Germany \\ (all)}} & \textbf{SVM-rbf} & \textbf{46.8$\pm$0.0} & \textbf{8.0} & 42.9$\pm$0.0 & 51.4$\pm$0.0 & 43.9$\pm$0.0 & 50.6$\pm$0.0 \\
          & MLP & 43.2$\pm$3.8 & 1.8 & 39.7$\pm$1.5 & 48.0$\pm$8.4 & 47.2$\pm$1.9 & 47.5$\pm$1.8 \\
          & FedAvg & 44.1$\pm$2.1 & 3.3 & 43.3$\pm$3.7 & 45.7$\pm$5.4 & 50.5$\pm$3.2 & 51.3$\pm$3.8 \\
    \bottomrule
    \end{tabular}
    \label{tab:multi_res}
\end{table}

\paragraph{\textbf{Multimodal Results}}
We evaluated the combined features of a multimodal dataset~\citep{buhler_detecting_2024} comprising four modalities: facial videos, eye tracking, and two physiological sensors (\autoref{tab:multi_res}). While the original authors reported their best performance using XGBoost with an $F_1$ score of 58\%, our best-performing model, SVM with RBF kernel, achieved a comparable $F_1$ score of 46.8 $\pm$ 0.0\%. Notably, unlike the authors, who found that combining all modalities yielded the highest performance, we observed that individual modalities performed similarly to the multimodal feature set. These differences may be attributed to the validation strategy, as the authors employed leave-one-person-out cross-validation, whereas we used a fixed data split and averaged results over five random seeds.

\begin{tcolorbox}[mynote,title={\textbf{Multimodal Results - Key Takeaway}}]
The performance differences between single-modality and multimodal feature sets highlight the need for additional multimodal datasets to enable more robust and generalizable conclusions. Furthermore, systematic feature optimization could potentially improve model performance across modalities.
\end{tcolorbox}

\subsubsection{Ablation Study}
In this ablation study, we investigate whether mind wandering can also be detected in the post-probe interval, after learners become aware of being off-task. While most prior work focuses on pre-probe signals, we hypothesize that post-probe segments may contain distinct behavioral and physiological patterns reflecting attentional recovery and re-engagement. From a cognitive perspective, this transition can be interpreted through the lens of meta-consciousness, defined as the explicit re-representation of one’s ongoing mental state \citep{schooler2002re}. When learners catch themselves mind wandering, a temporal dissociation occurs in which attention shifts from task-unrelated thought to meta-awareness of that state, potentially followed by translation dissociations if the original off-task experience is imperfectly reconstructed \citep{schooler2002re}.

\begin{figure}[t]
    \centering
    \includegraphics[width=10cm]{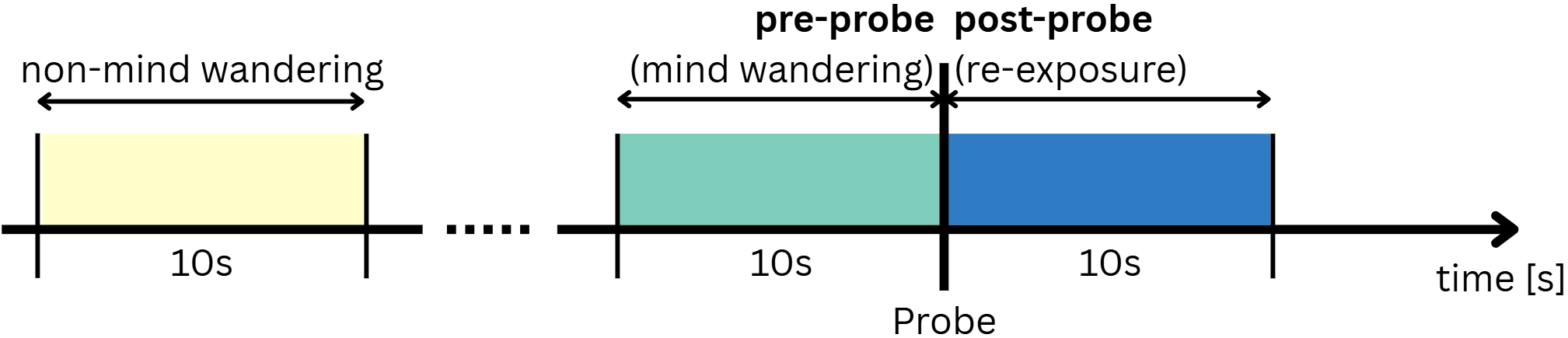}
    \caption{Example sample extraction in our main experiments and ablation study.}
    \label{fig:ablation}
\end{figure}

Reading studies indicate that participants frequently re-read missed content after reporting mind wandering, retracing about nine seconds of text, suggesting active cognitive reorientation~\citep{varao2017re}. Supporting this interpretation, empirical evidence indicates that re-reading occurs in up to 45\% of mind wandering episodes and is equally likely following both self-caught and probe-caught reports \citep{sousa2015re}. Such compensatory behaviors are most likely when learners perceive a need for clarification, underscoring re-reading as a deliberate re-engagement strategy rather than residual distraction \citep{sousa2015re}. Similarly, recent work in video-based learning~\citep{ebbert2025distraction} examines whether rewinding after off-task thoughts supports learning recovery. Building on these findings, we posit that post-probe periods may carry identifiable markers of such recovery behaviors, offering a new window into understanding how learners self-regulate and re-engage following mind wandering. In our experiments, we retained the standard 10-second non-mind wandering samples and added post-probe segments as positive samples (\autoref{fig:ablation}). Models trained on these post-probe samples were compared to the best pre-probe models using the same setup across three datasets: Germany multimodal (video domain), Korea, and our eye gaze datasets, where full recordings were available.

\begin{table}[t]
    \centering
    \scriptsize
    \caption{Comparison of the best-performing models across traditional ML, MLPs, and federated approaches under pre- and post-probe positive sampling conditions based on the mind wandering $F_1$ score, with the sampling strategy indicated in the sampling column.}
    \begin{tabular}{p{8mm} p{10mm} p{13mm} p{12mm} p{7mm} p{13mm} p{13mm} p{13mm} p{13mm}}
    \toprule    
    \raisebox{-0.5\height}{\textbf{Dataset}} & \raisebox{-0.5\height}{\textbf{Sampling}} & \raisebox{-0.5\height}{\textbf{Best model}} & \raisebox{-0.5\height}{$\mathbf{F_1}$ \textbf{[\%]}} & \raisebox{-0.5\height}{\textbf{AC [\%]}} & \raisebox{-0.5\height}{\textbf{Prec. [\%]}} &\raisebox{-0.5\height}{\textbf{Rec.[\%]}} &  \raisebox{-0.5\height}{\textbf{AUC [\%]}} & \raisebox{-0.5\height}{\textbf{Acc. [\%]}}\\
    \\
    \midrule
    Germany & before & SVM-lin & 61.2$\pm$0.0 & 32.9  & 52.0 $\pm$ 0.0 & 74.3 $\pm$ 0.0 & 33.2 $\pm$ 0.0  & 60.2 $\pm$ 0.0\\
    (video) & after & LR & \textbf{65.9$\pm$0.0} & 41.0 & 57.4 $\pm$ 0.0 & 77.1 $\pm$ 0.0 & 79.1 $\pm$ 0.0 & 66.3 $\pm$ 0.0  \\
     \midrule
    Korea & before & SVM-rbf & 42.6$\pm$0.0 & 23.3  & 35.4 $\pm$ 0.0 & 53.5 $\pm$ 0.0 & 55.5 $\pm$ 0.0  & 63.7 $\pm$ 0.0\\
    (video) & after & NB & \textbf{46.2$\pm$0.0} & 28.1 & 31.9 $\pm$ 0.0 & 83.7 $\pm$ 0.0 & 66.7 $\pm$ 0.0 & 50.9 $\pm$ 0.0  \\
     \midrule
    Ours & before & LR & 66.3$\pm$0.0 & 35.8  & 61.1 $\pm$ 0.0 & 72.4 $\pm$ 0.0 & 67.5 $\pm$ 0.0  & 64.8 $\pm$ 0.0\\
    (gaze) & after & GBoost & \textbf{77.3$\pm$0.9} & 47.0 & 67.5 $\pm$ 0.8 & 90.5 $\pm$ 1.0 & 76.7 $\pm$ 1.0 & 69.6 $\pm$ 1.2  \\
    \bottomrule
    \end{tabular}
    \label{tab:ablation}
\end{table}

Our results revealed that across all three datasets, models trained on post-probe samples achieved higher performance than those trained on standard pre-probe mind wandering detection data. In particular, the Germany dataset showed an improvement in $F_1$ score (from 61.2\% to 65.9\%), while our dataset demonstrated a substantial $F_1$ gain (from 66.3\% to 77.3\%). The Korea dataset also exhibited improved performance with post-probe training compared to pre-probe samples (from 42,6\% to 46.2\%), although absolute performance remained lower due to the small sample size (15 participants). These findings consistently suggest that post-probe samples capture detectable patterns related to material re-exposure or learners’ individual exploration of the just-presented content, which can serve as indicators of preceding mind wandering episodes. Overall, the results support our hypothesis that post-probe samples contain distinctive re-engagement signals, making them suitable for detecting prior mind wandering.

\begin{tcolorbox}[mynote,title={\textbf{Mind Wandering Detection through Post-Probes?}}]
Post-probe samples show strong potential for detecting previous mind wandering episodes by capturing distinctive re-engagement signals, though their use in real-time applications may introduce additional delays.
\end{tcolorbox}

\subsection{Discussion of the Modular Benchmarking}
Model performance variability across datasets highlights the difficulty of achieving real-world consistency, driven by differences in dataset size, data quality, recording conditions, participant diversity, and demographics. Importantly, four datasets \citep{lee_predicting_2022, hutt2024webcam, jaiyeola_one_2025, steadman2025difficulty} were collected in real-world settings, allowing us to partially capture ecological variability and assess model robustness beyond controlled laboratory environments. While this strengthens the practical relevance of our findings, it also underscores the need for more large-scale, ecologically valid datasets that reflect diverse learning contexts. Such data are crucial for developing models that generalize reliably and can be deployed in authentic educational applications.

Furthermore, the use of facial videos and other physiological signals raises important ethical and legal concerns, particularly in light of evolving regulations such as the AI Act \citep{AIAct}. These data modalities are highly sensitive, as they may reveal biometric identifiers, health-related information, or affective states beyond the original learning context. Future research must therefore prioritize privacy-preserving approaches, such as federated learning, to avoid centralized storage and sharing of raw user data. In particular, feature-based mind wandering detection combined with federated learning, where raw signals remain on-device and only latent feature representations (e.g., hidden, previously extracted feature vectors or facial landmark coordinates derived through local preprocessing) are used for model updates, can substantially reduce privacy risks. Under such settings, the original signals cannot be directly reconstructed from the shared representations, limiting the feasibility of information leakage even under advanced attacks or model unlearning scenarios. Empirical evidence from our benchmarks further indicates that federated learning can achieve performance comparable to centralized approaches, supporting its adoption as a practical and effective strategy for privacy-preserving mind wandering detection \citep{bodonhelyi2026safeguarding}. In addition, strict data minimization, secure storage, encryption, and transparent consent procedures are essential to ensure that learners retain control over how their data are collected, processed, and reused. Addressing these concerns is critical for building trust and enabling the responsible deployment of mind wandering detection systems in real-world educational settings.

Benchmarking mind wandering detection reveals both the efficacy and constraints of current approaches across modalities. A consistent finding is that traditional ML models frequently surpass or match complex deep learning (DL) architectures in performance across EEG, eye-tracking, and physiological signals. This may be attributed to the small size and heterogeneous nature of existing datasets, which curtails the typical advantages of DL. Specifically, in the video modality, simple models (e.g., MLPs, decision trees) proved competitive, underscoring the value of computational efficiency and interpretability in educational contexts. The robust performance of MLPs and logistic regression in eye-tracking, independent of task domain, suggests the generalizability of these attentional features. While logistic regression showed promise with physiological signals, data scarcity limits definitive conclusions. Furthermore, multimodal analyses demonstrated only marginal gains over single-modality inputs, indicating that enhanced predictive performance is not guaranteed without systematic feature integration and optimization. This highlights the critical need for developing harmonized multimodal datasets to establish the conditions under which integration is maximally beneficial.

The ablation study provides preliminary evidence that post-probe intervals contain distinctive re-engagement signals valuable for mind wandering detection. While most prior research focuses on pre-probe detection, our findings suggest that post-probe behavioral and physiological dynamics also hold predictive value. Given the limited related literature, future work should prioritize examining post-probe detection in adaptive scenarios to yield more ecologically valid signals of attentional shifts. Moreover, expanding the investigation across diverse educational activities and multiple modalities is essential to clarify the generality of these post-probe markers.

Previous research increasingly links mind wandering to underlying neurodevelopmental and personality-related factors, including ADHD-related hyperactivity–impulsivity. Large-scale studies show that mind wandering mediates the effects of neuroticism on insomnia \citep{zhang2023influence} and the relationship between internet addiction, anxiety, and hyperactivity–impulsivity in adolescents \citep{li2025role}, highlighting its role as a transdiagnostic marker of self-regulation difficulties. These findings suggest that mind wandering may reflect stable cognitive vulnerabilities rather than purely situational lapses, underscoring the need for more datasets including neurodivergent learners to enable personalized models that generalize reliably within specific neurodivergent groups.

\paragraph{\textbf{Potential real-world applications}} Translating mind wandering detection into real-time online learning applications is constrained by several factors. Effective deployment requires low-latency processing for timely feedback, yet many high-accuracy models, particularly deep learning architectures, incur computational overhead that compromises utility. Furthermore, current false positive rates risk unnecessary interruption, potentially causing learner frustration and disengagement. While EEG offers strong predictive potential, its cost, complexity, and intrusiveness render it impractical for home-based educational use. The most scalable and accessible approach involves integrated or low-cost webcams for facial video analysis and eye-tracking, despite generally yielding lower predictive performance. Alternative physiological monitoring via wearable wristbands is possible but requires user purchase and adoption. When reliably implemented, such systems could enable targeted pedagogical support, including real-time refocusing prompts, post-hoc summaries of potentially missed content, and adaptive review questions that reinforce learning after lapses in attention. Rigorous empirical validation is essential to confirm whether integrating these interventions enhances learning outcomes. Future systems must therefore prioritize low-intrusion, cost-effective sensing, while incorporating personalization to support diverse learners, particularly neurodivergent individuals \citep{jaiyeola_one_2025}, ensuring both accurate detection and inclusive intervention design.

\subsection{Limitations}
One limitation of our study is that we did not apply nested cross-validation for model evaluation. Instead, we relied on fixed data splits to ensure comparability across modalities and models and used 5-fold cross-validation for hyperparameter tuning. Future work should explore these approaches to further validate and possibly enhance model performance. Another limitation of this work is that the choice of optimization metric during hyperparameter tuning may substantially influence model performance and comparability across datasets. However, as the codebase is fully modular, future studies can readily replicate or extend our experiments using alternative optimization metrics.

While our benchmarking focused on generalized detection due to the high volume of available data, the lack of neurodivergent and culturally diverse datasets remains a significant bottleneck. Moreover, only one of the 14 included datasets provides explicit comprehension measures, limiting the ability to directly relate machine learning–based mind wandering detection to learning outcomes. Future research should therefore prioritize more diverse, individual difference-aware data collection that also integrates robust measures of comprehension, enabling the transition from generalized detection models toward personalized and pedagogically meaningful systems.

The ablation study is limited by the scarcity of full session recordings, restricting post-probe analysis to only three datasets. A similar analysis for EEG was precluded by the unavailability of continuous data; future work should explore whether these findings generalize to EEG, given its sensitivity to cognitive states. Lastly, the theoretical basis for classifying post-probe samples as re-engagement requires empirical validation via corroborating behavioral or subjective data to strengthen interpretation.

\section{Conclusion}
In this work, we conducted a \textbf{systematic review of automatic mind wandering detection within educational contexts}, identifying recent trends across learning environments, participant groups, detection approaches, evaluation strategies, and the use of multimodal datasets. Our analysis revealed several critical gaps in the literature, particularly the lack of standardized evaluation frameworks that enable comparison across studies. To address this, we developed and released an open, \textbf{modular benchmarking framework} that supports four commonly used modalities and can be flexibly adapted to both new and private datasets, thereby providing the first directly comparable overview across multiple mind wandering datasets. In addition, we contribute a \textbf{newly collected eye-tracking dataset} to support future research.

Our benchmarking study demonstrates that traditional machine learning models, particularly MLPs and logistic regression, frequently outperform more complex architectures across modalities, suggesting their practical suitability for deployment under varying computational and data constraints. Importantly, performance patterns varied systematically across learning tasks: video-watching tasks favored fast, lightweight models suitable for real-time detection, whereas reading-based tasks benefited from more stable representations where federated and centralized approaches performed comparably. These findings indicate that \textbf{task-aware model selection} is critical for practical application and that detection algorithms should be adapted to the cognitive and temporal characteristics of specific learning activities. Post-probe samples revealed distinctive signals of re-engagement that enhanced detection performance, but their delay constrains real-time applications. Using our framework, we conducted a comparative evaluation across 14 datasets and four modalities with more than 8000 experimental runs, offering the first unified benchmark for mind wandering detection in education and laying the groundwork for future personalization across tasks, learner populations, and educational contexts. Furthermore, substantial variability across datasets underscores the influence of recording environments, participant diversity, and cultural and learning-style differences, emphasizing the need for \textbf{more diverse, real-world datasets} to support robust and equitable deployment.

\section*{Declaration of generative AI and AI-assisted technologies in the writing process}
During the preparation of this work, the author(s) used ChatGPT and Gemini in order to increase the writing quality. After using this tool, the authors reviewed and edited the content as needed and take full responsibility for the content of the published article. 

\section*{Data Statement}
The dataset collected as part of this work is openly available on \href{https://zenodo.org/records/14097753?preview=1&token=eyJhbGciOiJIUzUxMiJ9.eyJpZCI6IjAyM2QxNWY1LWJlY2ItNDQ2ZC04YjNhLWUwYjdkZGU0MzU4ZCIsImRhdGEiOnt9LCJyYW5kb20iOiJkNTkyZmI0MzMxMWRiODVjNzJkNzhkNGYyOWEzMjMwYiJ9.MRLZkGqYIwafNVI-D9z1de1qY4o-iQ1fPJFK5NXgImrL095j3cWAoMhwjJHlbD3ZdeVWrmDuPH_ItHQY-wGMrQ}{Zenodo} under the Creative Commons Attribution 4.0 International License.

\section*{Declaration of Interest}
We have no conflicts of interest to disclose.

\bibliographystyle{elsarticle-harv} 
\bibliography{references}

\newpage
\appendix
\section{Summary of mind wandering detection modalities}
The following table provides an overview of the sensing modalities used for mind wandering detection and the corresponding ground truth collection methods reported in the literature. It highlights the methodological diversity across studies, including differences in probing strategies, self-reports, and retrospective measures.

\begin{table}[h]
    \centering
    \scriptsize
    \caption{Summary of mind wandering detection modalities and their associated ground truth collection methods. Multimodal and multi-dataset studies are listed in all relevant categories.}
    \begin{tabular}{p{8mm} c p{100mm}}
    \toprule
         \textbf{Modality} & \textbf{Collection} & \textbf{Published works} \\
    \midrule
          & self-caught & \cite{pain_msstnet_2025, china2023eeg, khan2022execute, khan2022feature, zhu2022topographynet, tasika2020framework, guo2019deep, grandchamp2014oculometric}\\
         EEG & probe-caught & \cite{asish_classification_2024, withammer-ekerhovd_classifying_2024, kuvar2023keystroke, MMSART2022eeg, dong2021detection, jin2020distinguishing, dhindsa2019individualized, jin2019predicting}\\
         & survey & \cite{riby_elevated_2025} \\
         \rowcolor[gray]{0.9}
          & self-caught & \cite{li_catching_2024, mills_eye-mind_2021, singha2021gaze, brishtel2020, faber2018automated, mills2016automatic, bixler2015automatic, hutt2024webcam}\\
          \rowcolor[gray]{0.9}
         eye-tracking & probe-caught & \cite{jaiyeola_one_2025, buhler_temporal_2025, asish_classification_2024, li_catching_2024, buhler_detecting_2024, rahnuma2024gazebased, asish_detecting_2022, MMSART2022eeg, lee_when_2021, bixler2021crossed, gwizdka2019exploring, hutt2017out, hutt2016eyes, bixler2016automatic, bixler2015, bixler2014toward, UZZAMAN20111882}\\
         skin & self-caught & \cite{brishtel2020}\\
         sensors& probe-caught & \cite{buhler_detecting_2024, MMSART2022eeg, chang2021efficient, bixler2015} \\
         \rowcolor[gray]{0.9}
         cardio. & probe-caught & \citep{buhler_detecting_2024, MMSART2022eeg}\\
         facial & self-caught & \cite{buhler2024lab, colorado2021, stewart2017} \\
         video & probe-caught & \cite{buhler_detecting_2024, buhler2024lab, lee2022predicting}\\
        \rowcolor[gray]{0.9}
         keyboard/ & self-caught & \cite{kuvar2023keystroke} \\
         \rowcolor[gray]{0.9}
         mouse & probe-caught & \cite{dias_da_silva_wandering_2020}\\
         fNIRS & correct & \cite{liu2021fnirsbased}\\
    \bottomrule
    \end{tabular}
    \label{tab:modality_gt_summary}
\end{table}

\section{Search String}
The following search string was utilized, presented in a generalized format rather than tailored specifically for Scopus or Web of Science. It illustrates the search terms and their logical connections.

\begin{lstlisting}[caption=Scopus Search Query, label=lst:search_query]
("mind-wandering" OR "cognitive drift" OR "mental drift" OR "attention lapse" OR "cognitive distraction" OR "mental diversion" OR "thought wandering" OR "cognitive wandering" OR "attention drift" OR "mind drifting" OR "loss of focus" OR "TUT" OR "task unrelated thought" OR "off task thought"  OR "stimulus independent thought" OR "unguided thought" OR "meandering thought" OR "unintentional thought" OR "spontaneous thought*" OR "daydreaming" OR  "internal attention" OR "off-task thought" OR "internally-directed attention")
AND
("machine learning" OR "ML" OR "DL" OR "deep learning" OR "artificial intelligence" OR "AI" OR "automatic" OR "neural network" OR "deep neural network" OR "NN" OR "support vector machine" OR "kNN" OR "nearest neighbours" OR "convolution*" OR "CNN" OR "ResNet*" OR "SVM" OR "classifi*" OR "regression" OR "recurrent"  OR "RNN" OR "Bayes*" OR "XGBoost" OR "gradient boosting" OR "LSTM" OR "BiLSTM"  OR "Long short term memory" OR "Locally weighted learning" OR "LWL" OR  "decision tree" OR "random forest"   OR "transfer learning")
 AND 
("electroencephalography" OR "EEG" OR "electrocardiography" OR "ECG" OR "electroencephalogram" OR "eye movements" OR "electrophysiology" OR "eye gaze" OR "eye" OR "gaze" OR "eye tracking" OR "functional magnetic resonance imaging" OR "fMRI" OR "functional near-infrared spectroscopy" OR "fNIRS" OR "galvanic skin response" OR "GRS" OR "electrodermal activity" OR "EDA" OR "video" OR "facial expression" OR "facial emotion recognition" OR "FER" OR "keystroke" OR "mouse dynamics" OR "mouse" OR "speech" OR "voice" OR "body" OR "posture" OR "log data" OR "context feature" OR "pupillometry" OR "heart rate" OR "multimodal" OR "gaze signal" OR "facial action unit" OR "area of interest" OR "AOI")
\end{lstlisting}

\section{Evaluation metrics}\label{app:ac_metric}
In this section, we describe how the above chance level metric used in this study were calculated~\citep{buhler2024lab}. The \textit{Above Chance Level} (AC) quantifies the improvement of the model over random or majority-class guessing. It is computed as follows:
\[
\text{AC} = \frac{ActualPerformance - Chance}{PerfectPerformance - Chance}
\]

\section{Overview of Machine Learning and Neural Network Algorithms} \label{app:algorithms}

\subsection{Traditional Classifiers}
\textbf{Support Vector Machines (SVMs)} \citep{cortes1995support} were evaluated using both linear and radial basis function (RBF) kernels. Linear SVMs are well suited for approximately linearly separable data, offer strong generalization through margin maximization, require few hyperparameters, and are computationally efficient. In contrast, RBF-kernel SVMs can model complex, non-linear decision boundaries by implicitly projecting data into a higher-dimensional feature space, but are more sensitive to hyperparameter selection and computationally more demanding. Overall, SVMs provide a robust trade-off between expressiveness and generalization, particularly in high-dimensional settings, but may scale poorly to large datasets.

A \textbf{Decision Tree} \citep{breiman_classification_2017} is a supervised learning algorithm that recursively partitions the feature space using axis-aligned splits selected to maximize class purity (e.g., Gini impurity or information gain). Decision trees are highly interpretable and can model non-linear relationships without requiring feature scaling. However, they are prone to overfitting, sensitive to noise and small data variations, and often exhibit limited generalization.

\textbf{Logistic Regression} \citep{hosmer2013applied} is a linear classification model that estimates class probabilities by applying the logistic function to a linear combination of features. It is computationally efficient and offers strong interpretability through feature weights. While regularization ($L_1, L_2$) prevents overfitting in high-dimensional settings and class weighting addresses imbalance, the model’s linear decision boundary may limit performance on complex, non-linear patterns without explicit feature engineering or basis expansions.

\textbf{Naive Bayes} \citep{zhang_optimality_nodate}, including \textbf{Gaussian Naive Bayes}, comprises probabilistic classifiers based on Bayes’ theorem with the assumption of conditional feature independence. These models are computationally efficient, robust to high-dimensional data, and perform well with limited training samples. However, their strong independence and distributional assumptions (e.g., normality in the Gaussian variant) constrain performance when features are correlated or exhibit non-linear dependencies.

\textbf{k-NN} \citep{cover_nearest_1967} is a non-parametric, instance-based algorithm that predicts labels based on the majority class among the 
\textit{k} nearest neighbors under a chosen distance metric. It makes minimal assumptions about data distributions and can capture complex decision boundaries. However, its performance is sensitive to the choice of
\textit{k}, distance metric, and feature scaling, and it is computationally expensive at inference time, particularly for large datasets.

\textbf{Ensemble models} combine multiple weak learners to improve predictive performance and robustness. \textbf{Random forests}~\citep{breiman2001random} construct an ensemble of decision trees using bootstrap sampling and random feature selection, with predictions aggregated via majority voting. This approach reduces variance and overfitting and performs well with minimal tuning, but can be less interpretable than single trees and may struggle with highly sparse or high-dimensional feature spaces.

\textbf{XGBoost} is a scalable, efficient implementation of gradient boosting that builds an ensemble of decision trees sequentially. Each new tree is trained to correct the residual errors of the combined ensemble so far, optimizing a differentiable loss function using gradient descent. Its high performance and flexibility make it well-suited for structured data and widely used in predictive modeling tasks. We also tested \textbf{Gradient Boosting}, a similar algorithm implemented using scikit-learn.

\textbf{Boosting-based ensemble methods}, including \textbf{AdaBoost} \citep{hastie_multi-class_nodate}, \textbf{Gradient Boosting} \citep{friedman2001greedy}, and \textbf{XGBoost} \citep{chen2016xgboost}, construct ensembles of weak learners—typically shallow decision trees—in a sequential manner. AdaBoost reweights training samples to emphasize previously misclassified instances and combines learners through a weighted vote, effectively reducing bias but remaining sensitive to noise and outliers. Gradient Boosting and XGBoost instead optimize a differentiable loss function by fitting successive trees to residual errors, enabling strong performance on complex, structured data. While these methods can model non-linear relationships effectively, they are computationally intensive and sensitive to hyperparameter choices, particularly the learning rate and number of estimators, which together control the bias–variance trade-off and risk of overfitting.

\subsection{Deep Learning}
MLP (Multilayer Perceptron) \citep{rumelhart_learning_1986} is a foundational feed-forward neural network consisting of stacked fully-connected layers with non-linear activation functions. By utilizing backpropagation and stochastic gradient descent to optimize a differentiable loss function, MLPs can capture complex, non-linear relationships within structured data. While MLPs offer immense representational power and architectural flexibility, they are highly sensitive to hyperparameter configurations, such as learning rate, batch size, and network depth, and generally require larger datasets than traditional ensemble methods to generalize effectively without overfitting.

\subsection{Federated Learning}
Federated learning shifts the paradigm of machine learning by training models locally on client devices to maintain data privacy, utilizing algorithms like FedAvg \citep{McMahan2017a} to aggregate these updates via weighted parameter averaging. While FedAvg provides a simple and secure baseline, TurboSVM \citep{Wang2024} advances this architecture by incorporating SVM principles to optimize the global model. By focusing on the extraction and aggregation of informative support vectors rather than raw weights, TurboSVM employs class-wise embedding updates and spreadout regularization to sharpen decision boundaries. This strategic refinement not only improves model convergence and class separation but also significantly reduces communication overhead compared to traditional averaging methods.

\section{Eye Tracking Data Collection Procedure}\label{app:eye_dataset}
Participants were invited to a two-part study: firstly, to the eye tracking study on site, and secondly, to an online web-based retention test one week after participation. For the eye tracking experiment, participants were presented general study information and demographic information was collected. Afterward, participants completed the pre-video test, which, together with the subsequent post-test, was designed to assess their knowledge of the topic of derivatives. The test consisted of ten questions and was intended to represent an exam, by setting an overall time limit of 13 minutes and enabling skipping and going back to previous pages. Furthermore, the test was a combination of retention, comprehension, and transfer questions with different answer formats, encompassing single choice, multiple choice, and numeric input.

Subsequently, participants received the instructions for the video watching task, that consisted of three videos, during which they were instructed to report if they are mind wandering. The following (translated) instructions were presented: ``Your main task is to watch and understand an educational video. While you are watching the educational video, you will probably notice at some points that you are no longer following the content. This is called ``drifting off.'' If at any time during the video you find yourself drifting off, please indicate this by pressing the SPACEBAR. Pressing the spacebar does not stop the video, instead it continues playing normally. Please make sure that you press this key as close as possible to the time of the drift. Remember, there is no right or wrong way to mark this drift. You just have to make sure that you press the key only when you realize that you have drifted. Please place one finger on the spacebar during the videos, so that you are ready to press it in the according moments.''. These instructions follow an established method in the field \citep{faber2018driven, kopp2016mind} which encompasses task unrelated thoughts and task related inferences, thus defining Mind Wandering as not attentively following the video content. Mind Wandering was measured by using self-caught reports, making it possible to indicate instances that do not correspond to specific probes \citep{kopp2016mind}. This method was chosen instead of the probe-caught method to not interfere with eye tracking, as external probes would disturb the natural viewing of the videos.

The eye tracker was turned on after the instructions and participants subsequently performed the video watching task. A nine-point-calibration and validation of the eye tracker was performed before the first and in between videos. Participants then received the same test as before as a post-video test to measure the increase in knowledge through the video. We included an instruction manipulation
check, where written instructions prompted participants to refrain from choosing one of the presented multiple-choice options, but to click on the headline instead. This was intended to exclude participants that did not read instructions attentively. Lastly, we collected consent for data use, informed the participants of the study goals and collected information for compensation. The link for the web-based retention test was sent to participants on the seventh day after the eye tracking study. Participants were able to take the test at any time of the day, however, the access was denied as soon as the eighth day after the eye tracking study started. The test was the same as the pre- and post-test of the eye tracking experiment.

\end{document}